\documentclass[sigconf]{acmart}

\AtBeginDocument{%
  \providecommand\BibTeX{{%
    \normalfont B\kern-0.5em{\scshape i\kern-0.25em b}\kern-0.8em\TeX}}}

\setcopyright{none}
\makeatletter
\def\@copyrightspace{\relax}
\makeatother

\begin{document}

\title{Tactical Patterns for Grassroots Urban Repair}


\author{Sarah Cooney}
\email{cooneys@usc.edu}
\affiliation{%
 \institution{University of Southern California}
}

\author{Barath Raghavan}
\email{barath.raghavan@usc.edu}
\affiliation{%
 \institution{University of Southern California}
 }


\begin{abstract}
The process of revitalizing cities in the United States suffers from balky and unresponsive processes---de jure egalitarian but de facto controlled and mediated by city officials and powerful interests, not residents.  We argue that, instead, our goal should be to put city planning in the hands of the people, and to that end, give ordinary residents pattern-based planning tools to help them redesign (and repair) their urban surrounds. Through this, residents can explore many disparate ideas, try them, and, if successful, replicate them, enabling bottom-up city planning through direct action.  We describe a prototype for such a tool that leverages classic patterns to enable city planning by residents, using case studies from Los Angeles as guides for both the problem and potential solution.
\end{abstract}


\begin{teaserfigure}
\centering
  \includegraphics[width=0.8\textwidth]{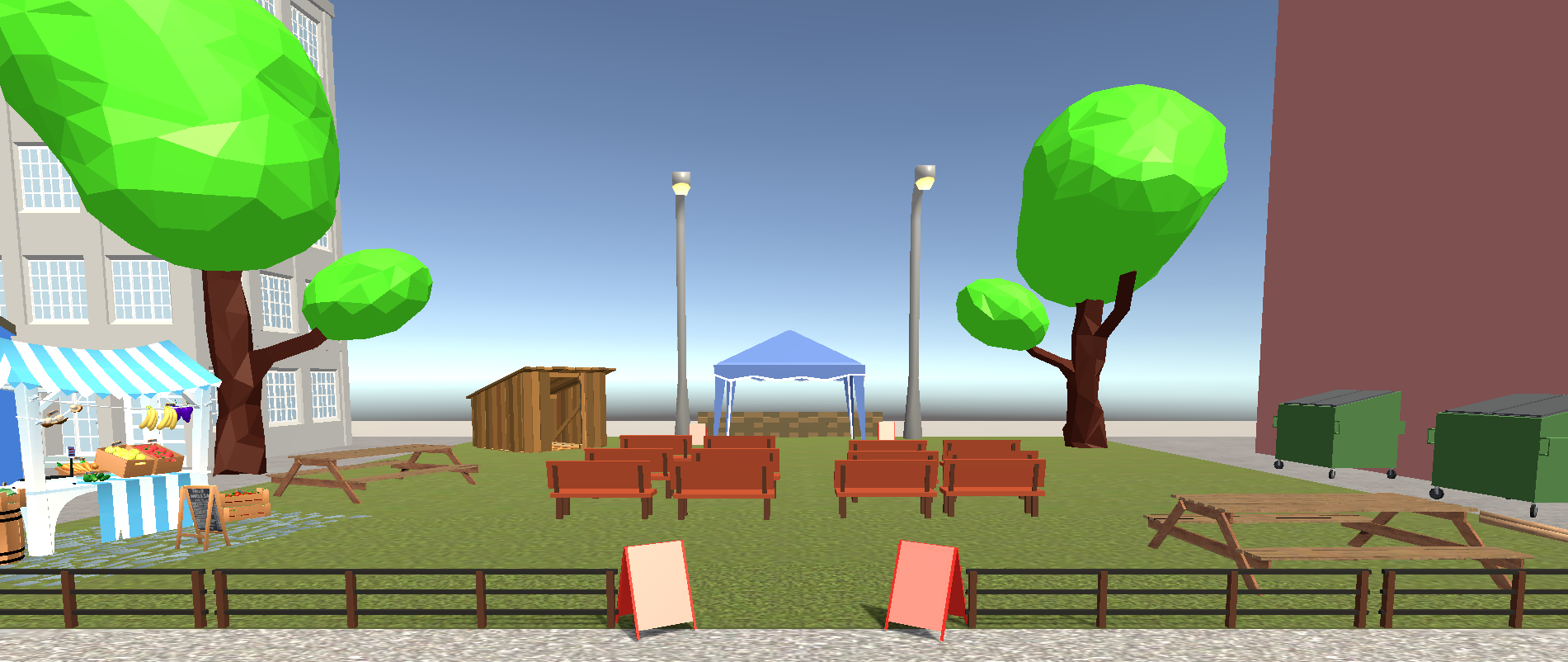}
  \caption{A design for a community theater space (Scenario A2) created with the PatternPainter tool.}
  \label{fig:teaser}
\end{teaserfigure}

\maketitle
\pagestyle{plain}

\section{Introduction}

\begin{quote}
{\em Even city officials seem to understand the importance of shade. Otherwise they probably wouldn't feel the need to bring canopies with them everywhere they go, as they did for both the groundbreaking for this project and last week’s ribbon cutting (below), right?

The shade gap is real. And the hotter our summers continue to be, the more uncomfortable it will continue to get for those that have the fewest choices about how they get around our streets. To not do better by those folks by taking their mobility needs into consideration in planning is, well...kind of shady.
}
\end{quote}
\noindent \emph{The Hoover Triangle: Effort to Do Bus Riders a Solid Takes Away their Shade}, Southern California Streets Initiative~\cite{streetsblog_17}

\begin{quote}
{\em The Bureau of Street Services (BSS) had hauled in four massive trees, and thick blue poles meant to anchor canopies had appeared...This was phase two, I was told.
Phase two?
Calling it phase two makes it sound like having to rip up concrete and rip out old new trees to install new new trees and new structures was always part of the plan...
Instead, the first ``phase'' had cost nearly two years of planning and \$600,000 in concrete, greenery, lighting, benches, and labor, all to yield profoundly underwhelming results...
At the groundbreaking, those involved in the project spoke of it as a model that could be replicated around the city and touted it as the fruits of what can happen when a community comes together to improve livability.
}
\end{quote}
\noindent \emph{New Shade Structures, Who Dis?: Hoover Triangle 3.0}, Southern California Streets Initiative~\cite{streetsblog_19}\\

Southern Californians are well aware of the fact that unshaded hardscape on a sunny day (i.e., almost every day) will absorb and then radiate heat, creating unbearable temperatures. A study by climate scientist Ariane Middel found that the temperature of unshaded asphalt was about 40 degrees Fahrenheit greater than its shaded counterpart~\cite{bloch2019shade}. With average summer high air temperatures of about 90 F (32 C), unshaded concrete temperatures can exceed 120 F (50 C), and in heat waves, such as that of Summer 2018 and Summer 2020, when \emph{air} temperatures in the region themselves exceeded 120 F (50 C), the lack of shade goes from uncomfortable to dangerous.

\begin{figure}[t!]
\includegraphics[width=0.40\linewidth]{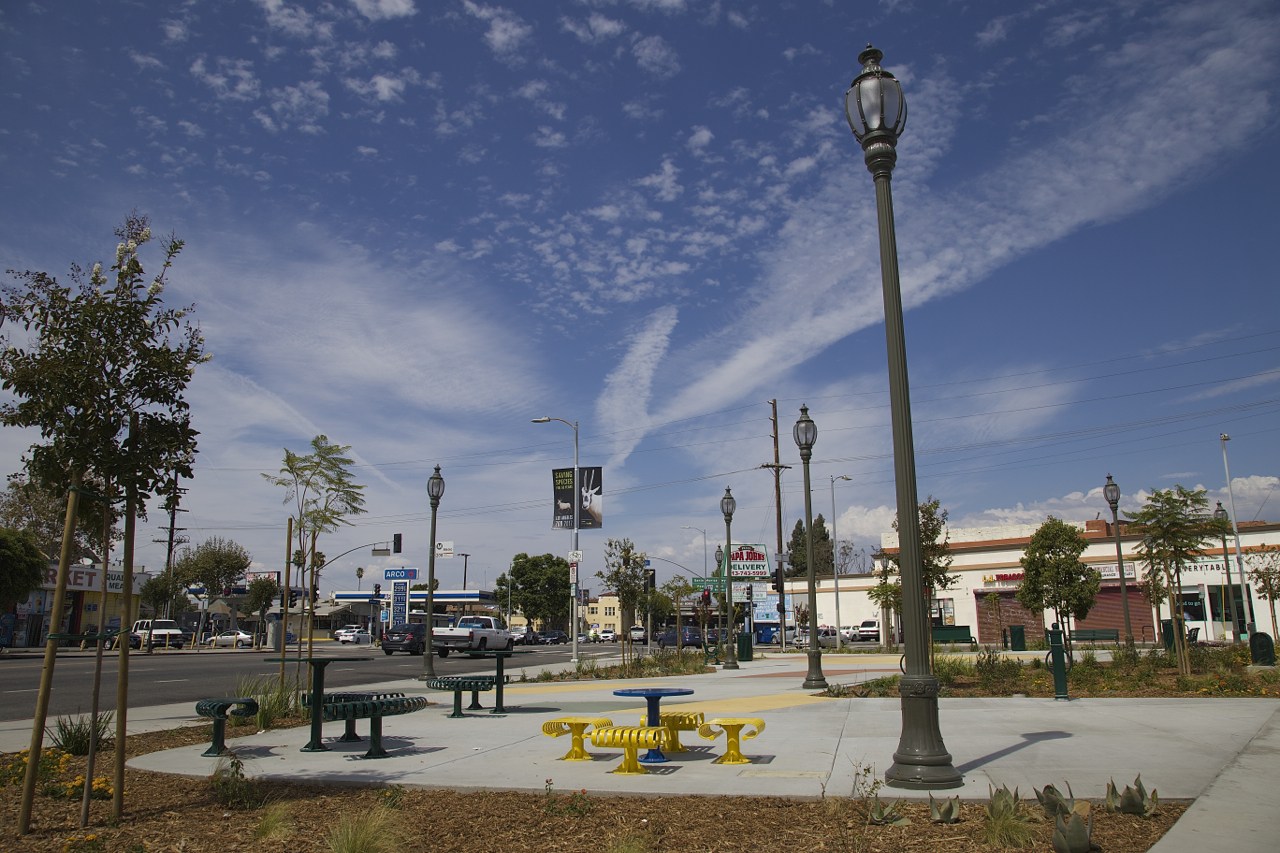}
\includegraphics[width=0.40\linewidth]{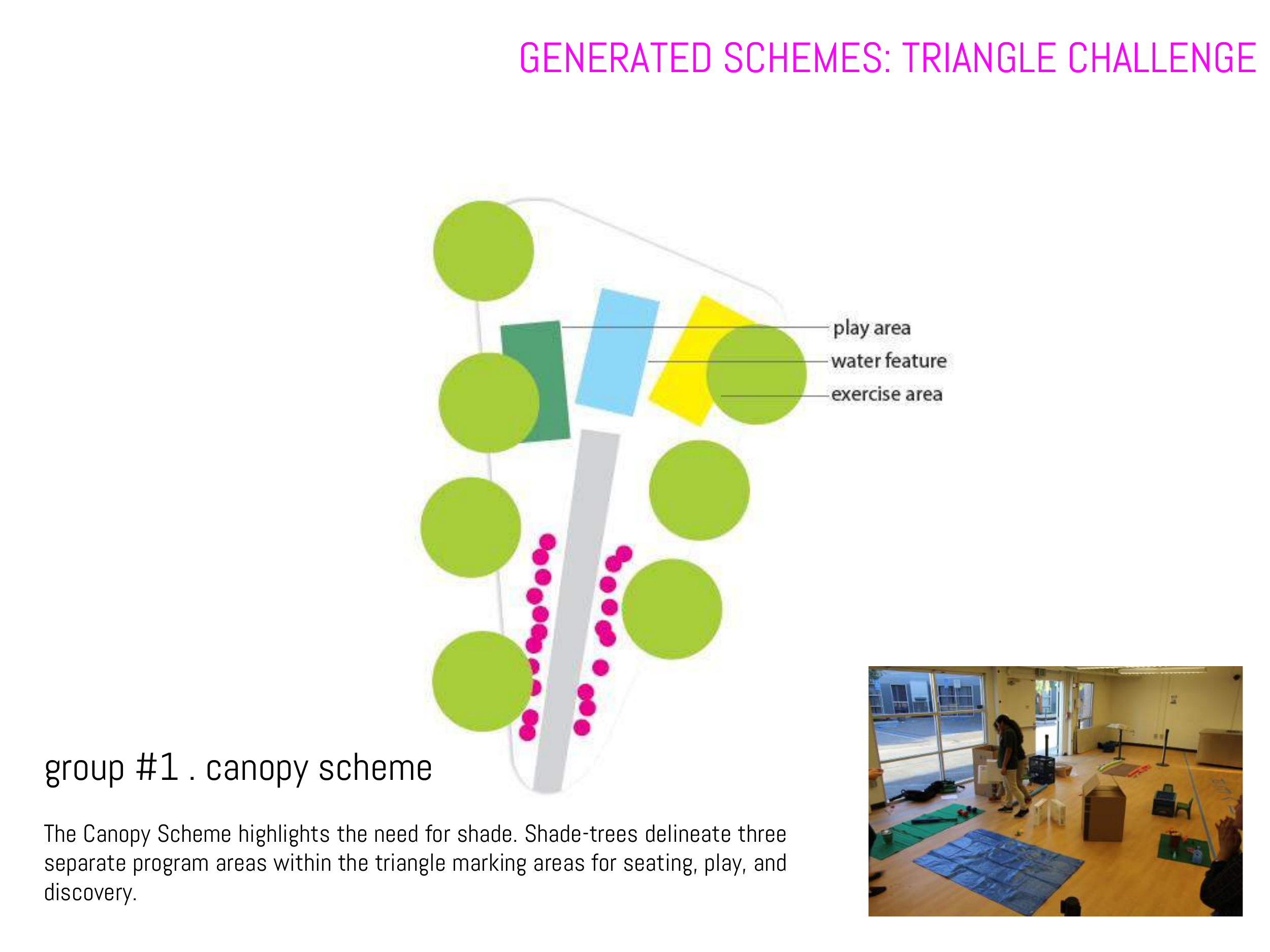}
\\
 \includegraphics[width=0.40\linewidth]{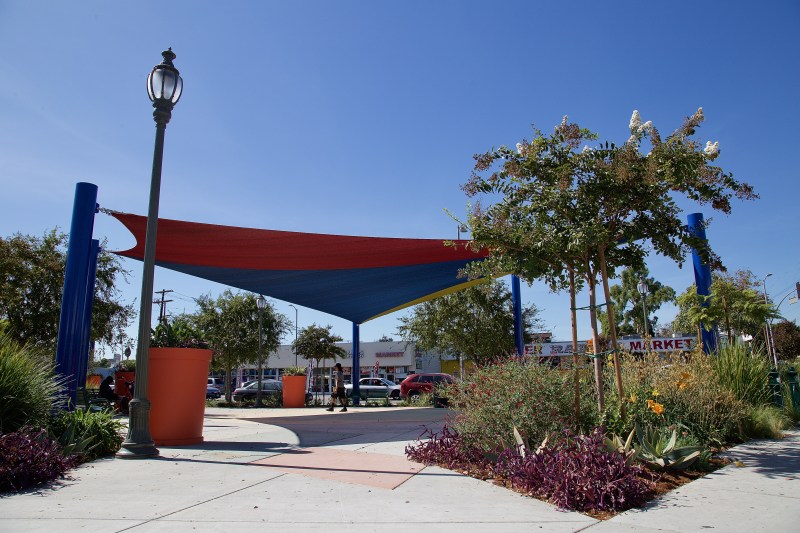}
  \includegraphics[width=0.40\linewidth]{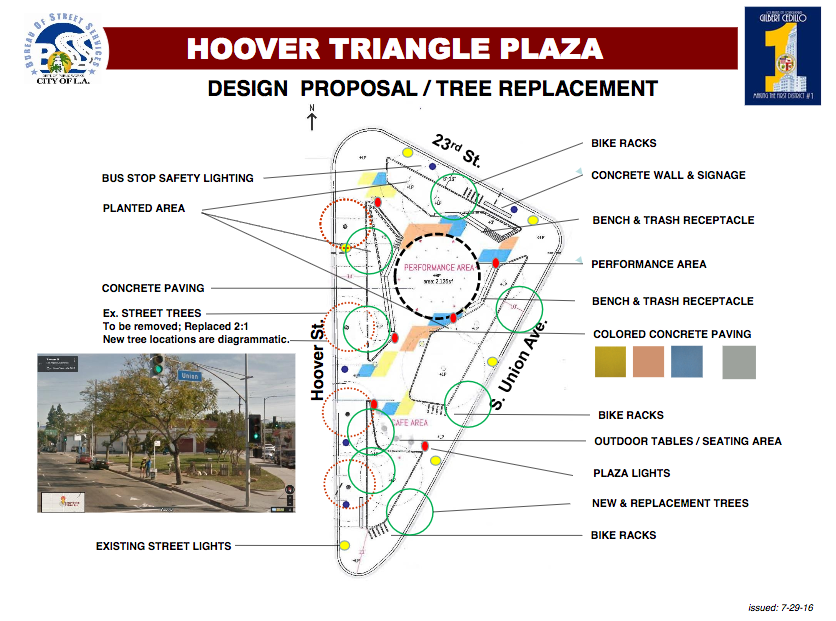}
    \caption{(From left to right) Top: The revamp of the Hoover Triangle with absolutely no shade \cite{streetsblog_17}; One of the plans for the Hoover Triangle made by workshop participants, featuring abundant shade (light green circles) \cite{meeting_record}. Bottom: The revamped revamp of the Hoover Triangle featuring ample shade \cite{streetsblog_19}; The official plans for the Hoover Triangle, featuring the removal of mature trees to be replaced by saplings unable to provide shade for at least a decade \cite{streetsblog_19}. }
   \label{fig:HooverCircle}
\end{figure}

Thus when asked to take part in participatory design workshops for the revitalization of the Hoover Triangle---a traffic island on Hoover street between 23rd and 24th streets in Los Angeles, an unloved piece of land home to two bus stops---community members wisely made shade their number one priority \cite{streetsblog_19}. Figure~\ref{fig:HooverCircle} shows one of the plans for the triangle generated by participants at a community workshop. The green circles indicate a desire for bountiful tree cover.  Other plans generated at the workshop also feature ample shade \cite{meeting_record}. 

However, the actual revamp---a product of 2 years of planning and \$600,000 in expenses---had no shade. Despite the added seating, lighting, and colorful concrete play areas, the plaza was essentially unusable without protection from the Southern California sun (see Figure~\ref{fig:HooverCircle}). So commenced `Phase 2', to revamp the revamp \cite{streetsblog_19}. Good intentions, and good processes and theories, are not good enough.  Indeed, in city revitalization, it is often the case that good processes, such as participatory design, yield little beyond providing cover for the preordained decisions of city officials. 

Here we look to a different, bottom-up perspective, in which ordinary residents of a city take into their own hands, and minds, the task of repairing their urban environs.
Our goal is to give the average citizen more power to initiate and influence the planning process in community repair projects. While some citizens may already take on this type of project on their own (see the rise in ``tactical urbanism'' taking place worldwide~\cite{lydon:2015:tactical}), for others the burden of planning, funding, recruiting volunteers, gathering materials, and more is too high. We hope our tools can decrease these burdens allowing more people to get involved.\\[0.5ex]

\noindent \textbf{The problem of participatory methods.}
We believe that many participatory methods are egalitarian only in their process, but that such a narrowly circumscribed ``participation'' is insufficient to create egalitarian outcomes.  For example, in city planning processes we have witnessed firsthand, city officials dictate constraints to city planners \emph{before participatory planning begins}, especially regarding budgets, and these constraints are often not shared with residents. Such constraints and objectives create preordained outcomes that are then justified, ex post facto, by the planners.  For example, planners may be forced---due to budget, space, and other exogenous constraints---to select those community-member ideas that align best with official, hidden, a priori constraints.  In addition, residents who participate in such planning meetings are seldom true reflections of the resident population~\cite{einstein2019participates}, but are instead either those who wish to uphold the status quo or those with pet issues that they raise with city staff on a regular basis, and thus their perspectives add little new to the discussion or are otherwise tuned out.
We believe we should put the tools of change in the hands of those whose city it is---those who visit or play in a park should be the ones to directly plan (and revitalize) it, should they choose to.\\[0.5ex]
\noindent \textbf{Our aim.}
In this work, we tackle the burden of the planning phase. Even for the simplest projects, some level of vision and imagination are required.  During the participatory planning process, professional planners walk community members through exercises to elicit values and ideas during a series of workshops and then turn those ideas into a plan.  We aim to help citizens generate their own 3D visualizations of urban plans without the need for professional planners or lengthy workshops. 

While there has been past work in the HCI community in the context of urbanism and planning~\cite{biloria2018social,disalvo2007imaging,le2015planning,schroeter2012engaging,takeuchi2012clayvision,underkoffler1999urp,vainio2019towards}, it has not supplanted official processes. In fact, there has been a great deal of scholarship on augmenting and working within the conventional participatory design process for urban planning~\cite{chow2011multi,christodoulou2018information,maskell2018spokespeople,stephanidis2019seven}. The CSCW community has engaged with grassroots urban activism~\cite{ghoshal2019design}, but as far as we can find, it does not directly address the process of designing new community spaces within the urban environment.\\[0.5ex]
\noindent \textbf{Contributions.} Our work makes the following contributions. 1) We identify and analyze a new problem domain (tactical, grassroots urbanism) that can substantially benefit from HCI study, but has largely not been considered in past work. 2) We prototype \textit{PatternPainter}, a design aid for urban repair projects, to explore one direction of work in this new area. We use as an exemplar the scenario of designing an urban parklet (small park) in an abandoned lot, a common challenge in urban areas across the world. We bridge the urban planning/HCI divide and the expert/novice divides by leveraging the classic planning tome \textit{A Pattern Language}~\cite{alexander1977pattern}. Although written more than 40 years ago, the language of ``Common Land'', ``Pedestrian Streets'', ``Roof Gardens'', and the like are more relevant than ever.  PatternPainter allows 3D elements to be placed within a scene to visualize designs and patterns.  
3) Using a series of experiments performed on Amazon's Mechanical Turk, we evaluate PatternPainter's ability to help ordinary people communicate their intentions for revitalizing abandoned urban spaces. 
4) Reflecting on our experience with PatternPainter as well as commentary from a design expert, we develop three general implications for design of technological tools tackling the planning phase of tactical, urban revitalization projects. \\[0.5ex]
\noindent \textbf{Overview.} We first discuss in detail the guiding ideas behind our work---participatory planning, the tactical urbanism movement, and \textit{A Pattern Language}, and then review related work in HCI and CSCW and on using 3D visualizations for urban planning. In Section~\ref{PatternPainter} we discuss the design and implementation of the PatternPainter tool. We then evaluate the software using a series of Mechanical Turk experiments. Finally, we conclude with a discussion about areas for future work and investigation in this domain and describe three general design goals for the development of technological tools for tactical, grassroots, urban planning. 

\section{Background}

In this section, we provide an overview of three urban planning principles that guided our work---Participatory Planning, the Tactical Urbanism movement, and \textit{A Pattern Language} \cite{alexander1977pattern}. 

\subsection{Participatory Planning}

Methods for engaging citizens in the urban planning process have been used and studied for decades.  In theory, these methods are sound, reflecting all the best scholarship in the sociology of group demographics and communication. Indeed, the leading manuals for conducting participatory exercises place strong emphasis on democracy and the equality of laypeople and professional planners. For example, Bernie Jones's \textit{Neighborhood Planning: A Guide for Citizens and Planners} states unequivocally, "In the best of all possible worlds, both the professional and citizen planner would be using the guide together, as they jointly set about drafting a neighborhood plan... This book uses a democratic, participatory planning approach, and the planner working without the people has perhaps picked up the wrong book!" \cite{jones1990neighborhood}.  

However, even the most sincere and well-intentioned planner is often not enough to overcome the destruction wrought by bureaucracy and money. Pre-determined budgets, the impetus to seek only the profitable, the interests of powerful and wealthy stakeholders, and lack of adequate time for eliciting deep citizen participation, among other bureaucratic burdens, are antithetical to a truly democratic process \cite{radil2019rethinking}. Consider the case of the New York City Board of Estimates in the late 1950s, where the public hearings were held on Thursdays, with executive sessions (where the actual decision making occurred) held on Wednesdays \cite{jacobs2016death}.  To be clear, this is not a problem of the past. In the early 2010s, a low-income Chicago community put together a digital petition protesting the development of a new pawnshop (what would be the fifth within a few blocks) in their community. Despite knowledge of this digital dissent, officials stated that they did not see enough in-person opposition at the community meeting and approved the shop. The meeting in question was held at 10am on a Wednesday, a highly inconvenient time for most members of a low-income, working class community \cite{erete2017empowered}. 

In other instances, what appears at the outset to have been a successful participatory project was in hindsight more paternalistic than participative. In \textit{Radical Cities}, an analysis of social housing programs in South America, McGuirk highlights Rio's mid-1990s slum upgrading program, Favela-Barrio, as an example of just such a project \cite{mcguirk2014radical}.

This is not to say that participatory planning always fails. With enough time, money, and---most importantly---careful attention to community context, participatory projects can be quite successful.  The Raising Places program to develop community health plans centered around improving children's health is an excellent example of what a truly participatory design process can look like \cite{curbed_raising}.  However, in this case the designers were hired by a private foundation and given nine months and a grant of \$60,000 per community to complete the process. Most localities simply cannot and do not provide this kind of time or money for community projects. 

Given these failures of the participatory planning process, people have started circumventing the official channels, turning to a strategy referred to as tactical urbanism, which we describe to next. 

\subsection{Tactical Urbanism}
``Tactical Urbanism'' is a new term for an old concept.  Coined in 2011 by urban planners Mike Lydon and Anthony Garcia, tactical urbanism is ``an approach to neighborhood building and activation using short-term, low-cost, and scalable interventions and policies''~\cite{lydon:2015:tactical}. Indeed, this describes the way cities often originated---built and organized by the people to serve the needs of increasingly complex societies. It was only with the invention of the modern state that cities were built top-down, according to comprehensive and organized plans, most frequently in grid-like formations \cite{scott1998seeing}. 

\begin{figure}[t]
\centering
\begin{minipage}{.48\textwidth}
  \centering
  \includegraphics[width=0.75\linewidth]{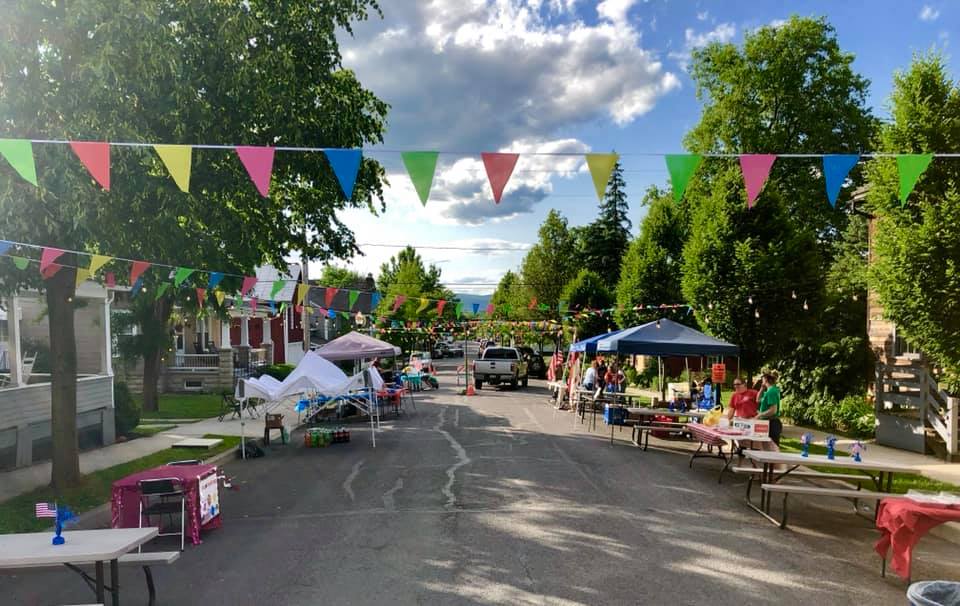}
  \captionof{figure}{An example of tactical urbanism: A pop-up street festival for Memorial Day 2019 held in Mifflinburg, Pennsylvania. Photo: David Cooney}~\label{fig:memDay}
\end{minipage}%
\hspace{1em}
\begin{minipage}{.48\textwidth}
  \centering
  \includegraphics[width=.55\linewidth]{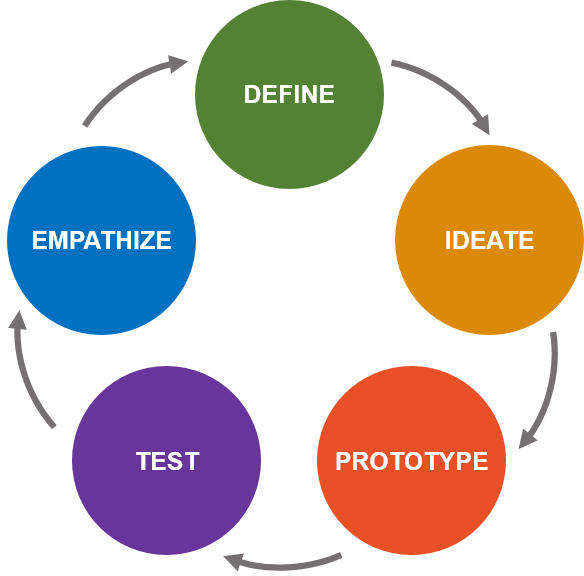}
  \captionof{figure}{The design thinking process, used to plan and execute tactical urbanism projects. }~\label{fig:designthinking}
\end{minipage}
\end{figure}

Examples of modern tactical urban projects include: Portland's intersection repair, in which intersections are painted to encourage slowed traffic and neighborhood cohesion; park(ing) day, an annual event during which parking spaces are turned into small parks; and pop-up shops, which promote the use of vacant buildings~\cite{parking,pop-up,cityrepair,tactical_vol1}. Figure~\ref{fig:memDay} shows the setup for a pop-up street festival in the town of Mifflinburg, Pennsylvania just before community members arrived to celebrate Memorial Day 2019. The festival featured food and children's games hosted by local community and school groups and even a temporary petting zoo on the lawn of a home owned by the local historical society.  

At its core, tactical urbanism is a reaction to the conventional city planning process: often politically fraught with a snail-like pace. ``For citizens, [tactical urbanism] allows the immediate reclamation, redesign, or reprogramming of public space''~\cite{lydon:2015:tactical}.  It is this last quote that truly captures our goal of putting the power for city building and urban repair directly into the hands of citizens. In reality, tactical urbanism exists on a spectrum from sanctioned projects---typically short events like LA's CicLAvia~\cite{ciclavia}, put on with the full cooperation of local authorities---to unsanctioned efforts like the lining of Durham's bike lanes with Jack O' Lanterns to make a statement about the city's failure to add protection for cyclists~\cite{streetsblog_bikes}. The perspective we take in this work is that while it is good that some residents have the support of their local authorities, we are not particularly concerned with doing things the ``official'' way, as we believe cities have, over human history, been organic entities and only recently have come to have regimented processes for their change and development~\cite{scott1998seeing}. 

To be clear, tactical urbanism is not advocating for complete anarchy or overthrowing city governments. Official channels are often necessary, for example, to complete and maintain large infrastructure projects or to provide services en masse such as comprehensive regional public transit. However, these large projects are often slow moving and expensive, and improvements to services often receive political pushback or are simply not prioritized. Tactical urbanism's role in interfacing with official channels is often as an activation mechanism in which temporary installations are used to highlight the potential for long-term change and to garner citizen support. Indeed, the well-loved pedestrian plaza that is New York City's Times Square began as a tactical demonstration with the placement of a few cheap folding chairs. Another example of a tactical project that spurred long term change is Matt Tomaluso's guerilla sign-posting campaign ``Walk Raleigh," which was not only adopted by Raleigh, but became the ``Walk [Your City]" campaign as it was adopted by numerous other municipalities across the country \cite{lydon:2015:tactical}.  We feel that the HCI community is already poised to bring expertise and solutions to the tactical urbanism movement due to past work in the areas of making, hacking, building, and repair \cite{bardzell2014making,rosner2014making,toombs2015proper} 

Lydon and Garcia adopt the design thinking process, developed by Tom and David Kelly (founders of the global design firm IDEO \cite{ideo_design}), to frame the process of tactical urbanism. The five steps are: empathize, who (both directly and indirectly) is affected by the problem or would be affected by a solution; define, identify the particular site of interest and clearly express the causes of the problem; ideate, develop methods or plans for addressing the problem at hand; prototype, plan a low-cost and quick to implement solution to the problem; and test, carry out the intervention and gather data and feedback~\cite{lydon:2015:tactical}.

In this paper we focus on the ideation phase, but believe that all phases are important. In the future, we aim to develop technical tools to assist in the other phases as well. Though we are focused on the ideation phase, it must be rooted in the findings of the empathy phase; ideation without empathy is likely to lead to the same underwhelming (or potentially even harmful) results exhibited by the Hoover Triangle example. To that end, we have developed PatternPainter, with the goal of giving citizens the capacity to plan urban repair projects without the bureaucratic nightmare of the conventional process, but with enough guidance to make the process manageable rather than too overwhelming to begin. 

\subsection{\textit{A Pattern Language}}
The inside jacket of the late 1970s urban planning epic \textit{A Pattern Language} reads, ``At the core of these books is the idea that people should design for themselves their own houses, streets, and communities...it comes simply from the observation that most of the wonderful places of the world were not made by architects but by the people,'' aligning perfectly with our vision of a bottom-up approach to urban planning and repair~\cite{alexander1977pattern}. With 253 patterns starting from ``Independent Regions'' and ending with ``Things from your Life,'' the book considers a comprehensive language for building and planning from the regional level down to individual rooms. 
Using \textit{A Pattern Language} as the inspiration for PatternPainter enables us to convey this wisdom to the inexperienced planner, but also leaves significant room for customization. This strikes a key balance, as we do not want to force a specific aesthetic or vision on the user. 

Another benefit of \textit{A Pattern Language} is that it has a variety of patterns from tried-and-true to out-of-the-box. For example, it is well documented that trees (pattern 171) contribute immensely to the livability of a community \cite{astell2019association}. However, for lifelong urbanites the idea of animals (pattern 74) living outside of a zoo or farm might be unthinkable, and sleeping in public (pattern 94) is usually seen as something to eradicate rather than something to embrace.  

While the work (and the metaphor) has been considered extensively in the HCI literature~\cite{borchers2001patterns,bouz2011analyzing,crabtree2002pattern,fischer2011urban,griffiths2000pattern,kim2015motif,knowles2015models,knowles2016design,mahyar2018communitycrit,pan2013pattern,sauppe2014designing}, we found that it has seldom been applied in HCI within its original context of city planning and further, to our knowledge, never with the intent of the original authors of embodying a grassroots approach. Quercia et al. do mention several of Alexander's patterns in the context of the urban environment, but their use is confined to analyzing existing streetscape images, not for the design of new spaces~\cite{quercia2014aesthetic}. 

One work of particular note is the \textit{Liberating Voices} pattern language~\cite{schuler2008liberating} that builds upon the work of Alexander~\emph{et al.}. This language of 136 patterns is designed to inform the responsible use of modern information and communications systems to create equity and practice justice. This is in direct alignment with our goal of creating technological tools to help ordinary citizens imagine, design, and implement urban repair and community building projects in their neighborhoods.  Although it was not informed by this pattern language at its inception, PatternPainter is something of a manifestation of many of the patterns in the language including ``Citizen Access to Simulations,'' ``Civic Capabilities,'' and ``Voices of the Unheard.''

\section{Related Work}
We review related work in two sub-areas: 1) urban planning in HCI, and 2) the use of 3D visualization in urban planning applications, which has its basis in computer graphics techniques, but is studied and applied in a wide variety of fields.

\subsection{Urban Planning in HCI}
As mentioned previously, past HCI work in citizen-oriented urban planning has been largely confined to the conventional participatory design process~\cite{chow2011multi,christodoulou2018information,stephanidis2019seven}.  This includes the sub-space \emph{digital civics}, a ``cross-disciplinary area of research that seeking to understand the role that digital technologies can play in supporting relational models of service provision, organization and citizen empowerment... and the potential of such models to reconfigure power relations between citizens, communities and the state"~\cite{vlachokyriakos2016digital}. While scholars in digital civics have studied urban issues such as the trust divide between citizens and local government \cite{corbett2018going} and configuring new means of citizen participation in local planning \cite{le2015planning,saad2012long}, its main goal is to equalize power relations between officials and ordinary citizens, not to bypass these official channels.  

However, there is some work in CSCW and HCI that has begun to move toward citizen-oriented urban planning. The works of Vivacqua and Bicharra Garcia~\cite{vivacqua2018personal} and Sun~\cite{sun2017leveraging} leverage the community social capital built around the kind of shared community spaces PatternPainter seeks to empower people to build. Mahyar et al.'s CommunityCrit system takes a step outside the conventional by enabling citizens to voice concerns and opinions about community issues and projects via crowdsourcing technology instead of the typical in-person workshop, but the data from their system is still passed to the local government to ultimately make the decisions \cite{mahyar2018communitycrit}.
Sasao et al. have also made strides in the area of engagement outside of official workshops with the use of systems to engage people in microtasks for community upkeep and collaborative social activities in existing community spaces~\cite{sasao2015activity,sasao2015support}. Sasao and collaborators also work on the problem of vacant lots and buildings, but their system is confined to data gathering and geared to facilitating small upkeep tasks rather than complete overhaul of the space \cite{sasao2016supporting}. Another example is the BlockbyBlock system~\cite{meng2019collaborative}, which was created by a community member to allow neighbors to collect data on local code violations or instances of neglect such as overgrown lawns or trash left at abandoned properties, and then encourages them to take action to help their neighbors to mitigate these issues.  This is exactly the kind of grassroots activism we wish to encourage with PatternPainter, but we hope to encourage more comprehensive overhaul of spaces based on expert design principles, rather than individual upkeep tasks. 

The CSCW community has a history of engagement with grassroots activism in the urban context~\cite{ghoshal2019design}. For instance, the literature has engaged with grassroots activists fighting evictions and gentrification in Atlanta~\cite{asad2015illegitimate} and with issues surrounding food such as community food sharing~\cite{ganglbauer2014think} and urban foraging~\cite{disalvo2007imaging}. However, to our knowledge this work does not extend to the type of tactical, grassroots urban planning PatternPainter has been designed to facilitate.  

Finally, we discuss several past CSCW and HCI projects with similarities to PatternPainter and identify how they differ. Mosconi et al. study the Itialian social streets movement, which uses hyper-local Facebook groups to engage communities in offline activities~\cite{mosconi2017facebook}.  The primary difference is that these are not all placemaking projects, and have no integration of design expertise. Similarly, Manuel et al. review a UK initiative that encourages neighborhoods to create their own local plans~\cite{manuel2017participatory}.  They focus on the impact of storytelling by local communities in creating these plans, but like~\cite{mosconi2017facebook}, they do not integrate design expertise or 3D visualization. Slingerland et al. describe their work on a project in the Hague that used a series of workshops, focus groups, and other participatory methods that resulted in a series of eight design guidelines for urban playgrounds~\cite{slingerland2020exploring}.  However, these guidelines are not integrated into any technological visualization system like PatternPainter. Perhaps the guidelines in~\cite{slingerland2020exploring} could be integrated as expertise in future iterations of PatternPainter.

\subsection{3D Visualization in Urban Planning} \label{related-3D}

Computer simulation has been a part of urban planning for approximately half a century, beginning with computational models and basic data visualization \cite{kamnitzer1972computers}.  With the improvement of GIS and digital mapping technology, 2D map-based visualizations became part of the planner's toolkit \cite{yeh1999urban,zhang2009application}.  While at first confined only to experienced professionals, GIS tools have increasingly become part of the participatory design process \cite{talen2000bottom}.  However, these tools are still largely used within the context of traditional participatory design dynamics, for instance through expert presentation or facilitation of mapping exercises \cite{radil2019rethinking}, or as a method of crowdsourcing information to be viewed and used by city officials in the form of geoquestionnaires \cite{czepkiewicz2018geo}. 

In the last decade and a half, the sophistication of computer graphics techniques has ushered in an era of 3D visualization in urban planning.  (Note, while 3D GIS or city modeling is sometimes used to refer to physical models---see \cite{franic20093d,ghawana20133d,ramirez2017participatory}---we confine the following discussion to computer-based methods.)  

Today, 3D city modeling is widely used across a variety of domains and applications, many of which fall under the purview of urban planners and designers.  Biljecki et al. identified 29 application areas including estimation of energy demand, emergency response planning, archaeology, and park design (which we explore further)~\cite{biljecki2015applications}. However, when used in participatory design, these applications are still largely confined to the traditional participatory system. For instance, in \cite{lange2005combining} the authors create a virtual model of the proposed site for a set of wind turbines. They then hold a participatory workshop in which they facilitate the viewing and modification of the model to show various factors like how visible the turbines will be from certain locations and how much shadow they will create at different times of day. At the conclusion of the workshop, stakeholders offered feedback on the proposal that was incorporated into a revision.

Another trend in 3D city modeling is the use of city building games such as SimCity and Minecraft in planning education and participatory workshops \cite{blockbyblock,minnery2014toying}. One such game, Cities: Skylines has an extensive API that has been used to create realistic models of real cities \cite{cities_stockholm,cities_finland}. While we are inspired by the interface and capabilities of such games, there are a number of drawbacks that make actually using one of these games as a base infeasible for our project.  Despite recent research to make creating real city models easier in Cities: Skylines, expert knowledge is still required to format the GIS input on which the model is based and to manually fine-tune the model after data has been imported \cite{olszewski2020developing}. Basing our tool on an existing commercial game would also mean that users must own a copy of the game and understand how to play it.
By creating a simpler, web-based model we can host the tool cheaply and make it free for use.  It also allows us to make the entire tool open-source.\footnote{PatternPainter code can currently be obtained upon request, and will be publicly released soon.}

The use of augmented and virtual reality (AR and VR) in planning applications is another emerging trend. VR has been used to conduct virtual 'go-along' interviews, about part of Oulu, Finland without ever leaving the lab \cite{kostakos2019vr}.
AR has also been prototyped as a method for showing municipal plans in-situ \cite{bohoj2011public}. The UN Habitat project extended this to allow community members to see plans they had developed in participatory workshops in the actual location where interventions were proposed \cite{unar}.  While we are interested in this kind of visualization for future work, it goes beyond the scope of the current project.  

One of the applications identified by Biljecki et al. is park design \cite{biljecki2015applications}.  As with other applications, 3D visualization in park design has also largely been confined to the traditional system. For example, Lu and Fang provide an overview of methodology to create a park visualization using tools from the ArcGIS suite, a popular professional GIS software out of reach of most ordinary citizens due to prohibitive cost and complexity \cite{lu2014computer}. Similar to PatternPainter, Kwon et al. also create a 3D visualization tool centered on the vacant lot repair scenario \cite{kwon2019land}. However, their software is designed to be used within the conventional workshop process with use by expert facilitators rather than directly by citizens.

\section{PatternPainter} \label{PatternPainter}

In this section, we describe the development of our prototype tool, PatternPainter. We first reflect on our design choices, and then describe the implementation details.

\subsection{Design}
\begin{figure}[t]
\centering
\includegraphics[width=0.95\linewidth]{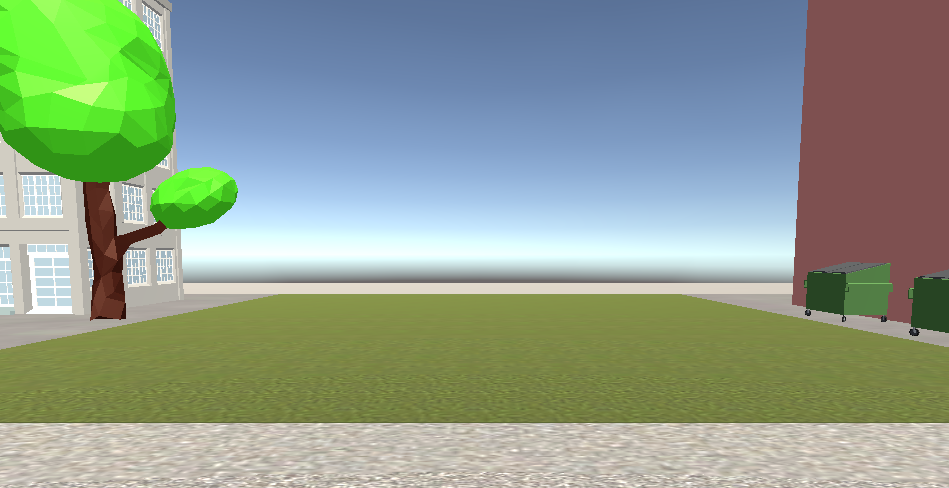}
    \caption{The scenario for PatternPainter: an empty urban lot ripe for repair.}
    \label{fig:emptyScene}
\end{figure}

As a case study for the prototype, we consider turning an abandoned lot into a small park (a parklet). This can be one of the simplest urban repair projects, but is known to have a statistically-significant impact on crime rates, mental health, and social connectedness of communities~\cite{heinze2018busy,moyer2019effect,south2018effect}.  Consider the resident who walks by the lot each day, who thinks it would be nice if the lot were cleaned up and turned into a community space, who might even participate in a cleanup if they knew how to begin.  But where to start? Just clean up the trash? Plant a few trees? Put in a bench or two? These are the questions PatternPainter aims to answer.  

The PatternPainter scene is set with an empty lot. The ``un-repaired'' lot is shown in Figure~\ref{fig:emptyScene}. For the initial model, the research team came up with 12 scenarios for repairing the lot inspired by some of Alexander's patterns that deal with uses for public space and community organization; for example, shopping street, accessible green, local sports, teenage society, and vegetable garden~\cite{alexander1977pattern}.  The full list of scenarios can be found in Table~\ref{Scenarios}.  We then developed a list of items that might be found in a space representing each scenario.  For instance, for scenario A4: \textit{The community would like to use this space for a community garden}, we included raised garden and flower beds, a utility shed, fences, goats, and chickens.  The item lists for each scenario were compiled and these items were added to PatternPainter as elements users can place in the scene. Figure~\ref{fig:teaser} shows a user design based on scenario A2: \textit{The community would like to turn this lot into an area where outdoor theater productions can be held during both the day and evening.} We would be remiss not to note that despite our best efforts to maintain neutrality of aesthetic, some of the design decisions do represent to some extent the aesthetic of the authors; see Section~\ref{results} for more discussion of this issue.

\subsection{Implementation}
PatternPainter was built using the Unity game engine \cite{unity}. The scene was created using a combination of public domain images and free assets and textures from the Unity Asset Store.  The 3D models and UI graphics are a combination of public domain images, free assets from the Unity Asset store, and free models downloaded from Sketchfab \cite{sketchfab}.\footnote{We will include attribution for the models and images in our public tool release.}  We chose to use Unity, which is freely available for non-commercial use, and source free models, as we wish the software to remain as accessible as possible.  

\begin{figure}[t]
\centering
\includegraphics[width=0.95\linewidth]{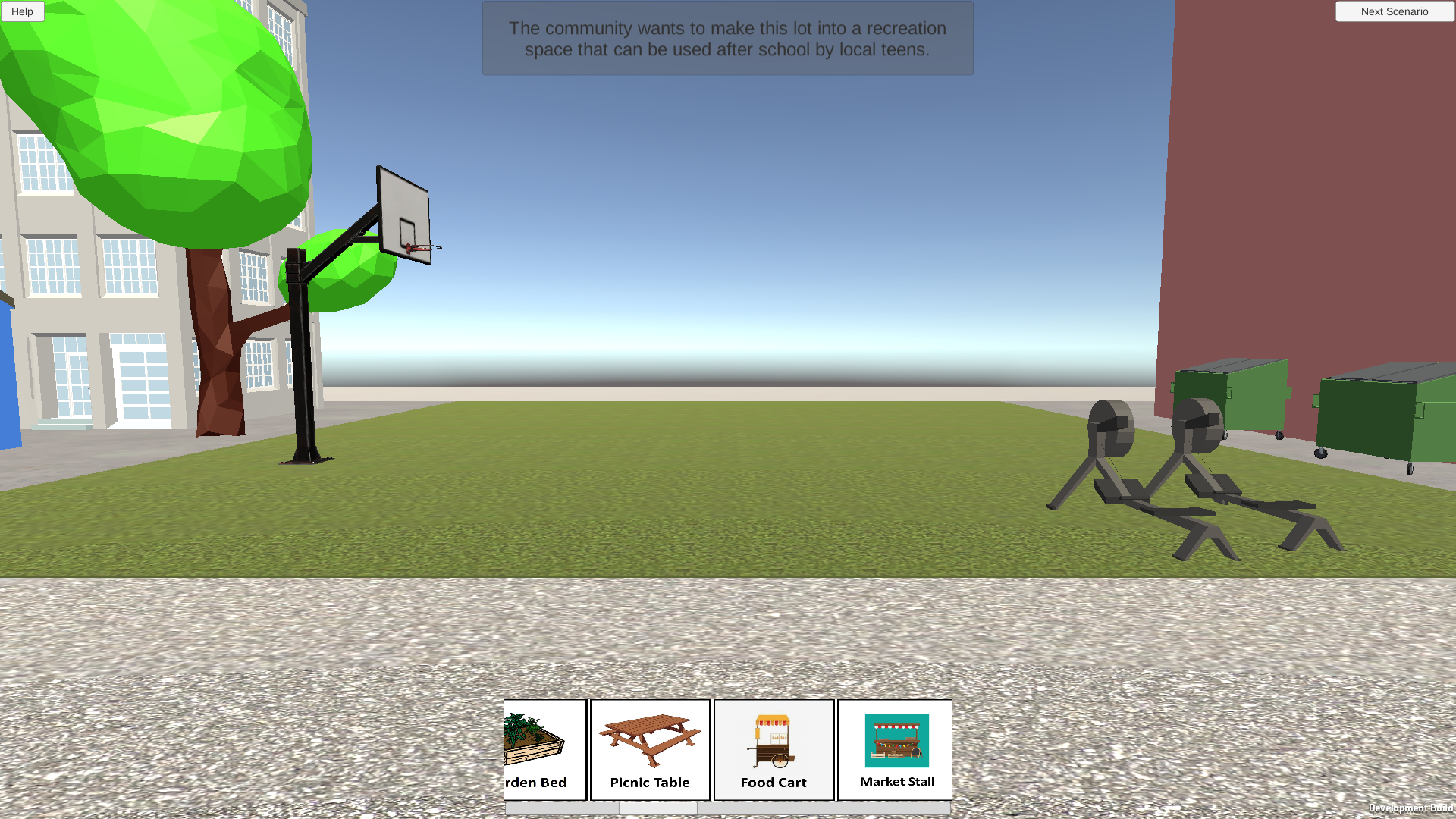}
    \caption{The PatternPainter user interface displaying scenario B2 with a partially completed design.}
    \label{fig:UI}
\end{figure}

The user interface, showing scenario B2, can be seen in Figure~\ref{fig:UI}. Game objects can be added to the scene using the the object menu located at the bottom of the interface and manipulated using a number of mouse and keyboard controls. The camera position and rotation can also be controlled with keyboard input. A help menu describing the various controls can be displayed by clicking the help button in the upper lefthand corner of the interface. 

The game was exported to javascript using the WebGL build feature in Unity and hosted on an AWS web server.\footnote{A fixed scenario can be tried here: \url{http://ec2-3-129-22-64.us-east-2.compute.amazonaws.com/BuildB}. Enter any text for the mechanical turk ID.}  Upon submission of each scenario, a screenshot of the scene is saved to the server. 

\section{Experiments}
The main goal of PatternPainter is to allow untrained individuals to effectively create designs and communicate goals for revitalizing their urban surrounds. Ideally, to validate that PatternPainter achieves this goal, we would partner with a community organization or neighborhood group performing an urban revitalization project and do an evaluation in the context of real-world use.  However, the COVID-19 pandemic has rendered this kind of evaluation temporarily infeasible, although we hope to be able to perform this kind of evaluation in the future. Therefore, in order to validate that PatternPainter achieves our goals, we used a series of online experiments performed using Amazon's Mechanical Turk. (For discussion regarding the ethics of using Mechanical Turk see Section~\ref{ethics}.)

\subsection{Experiment 1: Designs}
In the first experiment, participants used the tool to design community spaces based on the scenarios in Table~\ref{Scenarios}. Each participant was given one of the three sets of scenarios in randomized order.  For any considerations they might make regarding climate or weather, participants were instructed to assume the lot was located in Los Angeles, California, due to its fairly neutral year-round climate.  

\begin{figure}[t]
\centering
\includegraphics[width=0.95\linewidth]{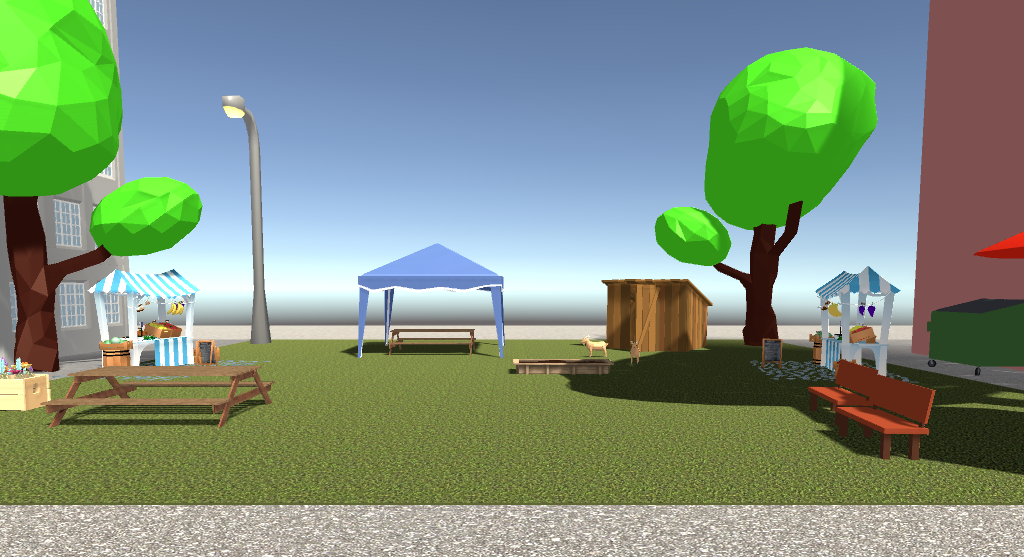}
    \caption{The practice scene, which participants in experiment one were asked to replicate.}
    \label{fig:practice}
\end{figure}

Before beginning the scenarios, for practice and validation, participants were asked to replicate the scene shown in Figure~\ref{fig:practice}. This ensured participants were familiar with adding and manipulating objects within the scene.  Participants who failed to replicate this test scene were rejected from the task. For this experiment we used participants who were located in the US and had achieved ``master'' status to ensure high-quality data.\footnote{In early trials where the master status was not required, we found people would simply leave a jumble of objects on the screen.  Due to the nature of online experiments, it was impossible to tell if it was a problem with the tool or if the workers simply were not making an effort to complete the task well.  We suspected the latter, but making such assumptions would have biased the data. Master status is conditional to continued review, and therefore incentivizes workers to take tasks more seriously.} The experiment was designed to take roughly half an hour, and participants were paid \$6.00 USD for completing a task successfully.  

\begin{table*}[]
\begin{tabular}{ll}
\multicolumn{1}{c}{\textbf{Group A}} & \multicolumn{1}{c}{}                                                                                                                                                                                                           \\ \hline
\multicolumn{1}{|l|}{1}              & \multicolumn{1}{l|}{The community wants a space where elderly residents can gather for leisure activities.}                                                                                                                    \\ \hline
\multicolumn{1}{|l|}{2}              & \multicolumn{1}{l|}{\begin{tabular}[c]{@{}l@{}}The community would like to turn this lot into an area where outdoor theater productions can be held during \\ both the day and evening.\end{tabular}}                          \\ \hline
\multicolumn{1}{|l|}{3}              & \multicolumn{1}{l|}{\begin{tabular}[c]{@{}l@{}}The community would like to see this lot across from the town hall transformed into a place where residents \\ and local leaders can meet one another informally.\end{tabular}} \\ \hline
\multicolumn{1}{|l|}{4}              & \multicolumn{1}{l|}{The community would like to use this space for a community garden.}                                                                                                                                        \\ \hline
\multicolumn{1}{c}{\textbf{Group B}} & \multicolumn{1}{c}{}                                                                                                                                                                                                           \\ \hline
\multicolumn{1}{|l|}{1}              & \multicolumn{1}{l|}{The community would like to see this area transformed into a space to hold a local farmers market.}                                                                                                        \\ \hline
\multicolumn{1}{|l|}{2}              & \multicolumn{1}{l|}{The community wants to make this lot into a recreation space that can be used after school by local teens.}                                                                                                \\ \hline
\multicolumn{1}{|l|}{3}              & \multicolumn{1}{l|}{The community wants to use this lot as a space where parents can take their children to promote healthy habits.}                                                                                           \\ \hline
\multicolumn{1}{|l|}{4}              & \multicolumn{1}{l|}{The community wants to turn the lot into an area where they can gather and host live music performances.}                                                                                                  \\ \hline
\multicolumn{1}{c}{\textbf{Group C}} &                                                                                                                                                                                                                                \\ \hline
\multicolumn{1}{|l|}{1}              & \multicolumn{1}{l|}{The community wants to turn this space into a park with plenty of shade and places to sit and relax.}                                                                                                      \\ \hline
\multicolumn{1}{|l|}{2}              & \multicolumn{1}{l|}{The community would like to see this lot turned into a park that local families can use with their children.}                                                                                              \\ \hline
\multicolumn{1}{|l|}{3}              & \multicolumn{1}{l|}{The community wants an after school location for children to study.}                                                                                                                                       \\ \hline
\multicolumn{1}{|l|}{4}              & \multicolumn{1}{l|}{The community would like to use the lot to set up a monument to their loved ones who passed away from accidents.}                                                                                          \\ \hline
\end{tabular}
\caption{The 12 scenarios used to implement and evaluate PatternPainter.}
\label{Scenarios}
\end{table*}

\subsection{Experiment 2: Validation}
The second experiment was used to evaluate the designs created in the first experiment, and determine how well users were able to communicate the intended uses for the space given in the scenarios using PatternPainter.  
Participants were told they were rating designs for revitalizing an abandoned lot in Los Angeles, California.  Participants were asked to rate designs on the eight metrics, listed in Table~\ref{tab:metrics}, on a scale of one to seven, as well as to briefly answer the following questions for each design: \textit{Please provide a brief description of how the community would use this space.  Who would use it?  What would they do? What is the purpose of the space?}. The practice designs and real designs were both given in randomized order, providing both quantitative and qualitative measures for evaluation.

Each survey consisted of 15 designs---three for practice, taken from initial trials of the experiment, which were the same across all surveys, and one design from each of the twelve scenarios.  The survey also contained four attention checks asking participants to choose a specific rating. Participants who failed two or more checks or entered nonsense text responses were rejected. Each survey was completed by five participants, meaning each design received five ratings.  Participants were restricted to users located in the US, but due to the ability to implement robust attention checks, were not restricted to master status. The survey was designed to take about 20 minutes, and participants were paid \$4.00 USD for each.  

In Section~\ref{results} we present and discuss the results of the experiments.

\begin{table}[]
\begin{tabular}{|l|p{2in}|}
\hline
\textbf{Metric}  & \textbf{Description}                                                                                                                          \\ \hline
Shade            & Are there shady spaces for people to spend time?                                                                                              \\ \hline
Play             & Are there activities available for children or young people?                                       \\ \hline
Comfort          & Are there places to sit and relax?                                                                                                            \\ \hline
Safety           & Are there places to supervise children playing, is there lightning for nighttime activities, etc.? \\ \hline
Access to Nature & Are there elements of nature such as trees, flowers, gardens, or animals?                          \\ \hline
Recreation       & Are there activities available for adults?                                                                                                    \\ \hline
Entertainment    & Could the area be used for performances, dancing, outdoor dining, etc.?                            \\ \hline
Sociability      & Would people enjoy gathering here to spend time with friends?                                      \\ \hline
\end{tabular}
\caption{The eight metrics used to evaluate the designs produced by PatternPainter.}
\label{tab:metrics}
\end{table}

\subsection{Ethical Considerations for Using Mechanical Turk} \label{ethics}
The research team feels we would be remiss not to acknowledge the ethical implications of experimentation using Amazon Mechanical Turk, given the precarious and often seriously underpaid nature of working on the platform~\cite{atlantic_turk}. As one participant noted via email, ``Rejections are very serious for workers,'' particularly those trying to maintain a master status. Another stated, ``Mturk is my only source of income at this time and I can't afford a rejection, which lowers my rejection score significantly and my chances of higher-paying surveys,'' driving home the precarious nature of this kind of work. We paid at a rate of \$12.00 USD per hour, based on the expected completion time, which is more than the minimum wage in a majority of US states and far higher than the federal minimum wage of \$7.25 an hour.  Despite our best efforts to be exceptionally clear in the instructions, there was some confusion about the validation task in our first experiment---in hindsight likely due to our use of the word ``practice'' instead of ``validation''---so participants who contacted us about this were given an opportunity to complete the validation and have the rejection reversed.

\section{Data and Results}
\label{results}
In this section, we present the results of the experiments, using both quantitative and qualitative analysis.

\subsection{Quantitative Analysis}
The quantitative data was gathered by asking participants to rate each design on the eight metrics found in Table~\ref{tab:metrics}. We had 28 different designs for each of the 12 scenarios. Each design received five sets of ratings for each metric, which were then averaged, resulting in 28 ratings for each metric for each scenario. These average design ratings were averaged to obtain a final rating on each metric for each of the 12 scenarios. Note that given the relatively small sample size we do not perform any significance testing.   

Three members of the research team independently chose what they believed to be the top three metrics representing each of the scenarios.  The top metrics, as shown in Table~\ref{topmetrics} were taken to be those that all three team members had in their top three, resulting in one or two top metrics per scenario.  
\begin{table}[]
\begin{tabular}{ll}
\hline
\multicolumn{1}{|l|}{\textbf{Scenario}} & \multicolumn{1}{l|}{\textbf{Top Metrics}}          \\ \hline
\multicolumn{1}{|l|}{A1}                & \multicolumn{1}{l|}{Comfort}                       \\ \hline
\multicolumn{1}{|l|}{A2}                & \multicolumn{1}{l|}{Entertainment}                 \\ \hline
\multicolumn{1}{|l|}{A3}                & \multicolumn{1}{l|}{Comfort, Sociability}          \\ \hline
\multicolumn{1}{|l|}{A4}                & \multicolumn{1}{l|}{Access to Nature, Sociability} \\ \hline
                                        &                                                    \\ \hline
\multicolumn{1}{|l|}{B1}                & \multicolumn{1}{l|}{Recreation, Sociability}       \\ \hline
\multicolumn{1}{|l|}{B2}                & \multicolumn{1}{l|}{Sociability}                   \\ \hline
\multicolumn{1}{|l|}{B3}                & \multicolumn{1}{l|}{Play, Safety}                  \\ \hline
\multicolumn{1}{|l|}{B4}                & \multicolumn{1}{l|}{Entertainment, Sociability}    \\ \hline
                                        &                                                    \\ \hline
\multicolumn{1}{|l|}{C1}                & \multicolumn{1}{l|}{Shade, Comfort}                \\ \hline
\multicolumn{1}{|l|}{C2}                & \multicolumn{1}{l|}{Play, Safety}                  \\ \hline
\multicolumn{1}{|l|}{C3}                & \multicolumn{1}{l|}{Comfort, Safety, Sociability*}                             \\ \hline
\multicolumn{1}{|l|}{C4}                & \multicolumn{1}{l|}{Comfort, Access to Nature}     \\ \hline
\end{tabular}
\caption{The top metrics representing each scenario as determined by the research team. 
\\ *For scenario C3, there was no metric agreed upon by all three members of the research team. The metrics given were agreed upon by two of the three members.}
\label{topmetrics}
\end{table}

Table~\ref{tab:averages} shows the average for each metric for each scenario.  For each scenario, the metric with the highest average is given in bold, while the metrics chosen as most representative for each scenario (see Table~\ref{topmetrics}) are given in italics.  

\begin{table*}[]
\begin{tabular}{|l|l|l|l|l|l|l|l|l|l|}
\hline
  & Example Design & Shade  & Play   & Comfort  
   & Safety & Access to Nature   & Recreation   & Entertainment & Sociability            \\ \hline
A1 & 
    \includegraphics[width=1.25in]{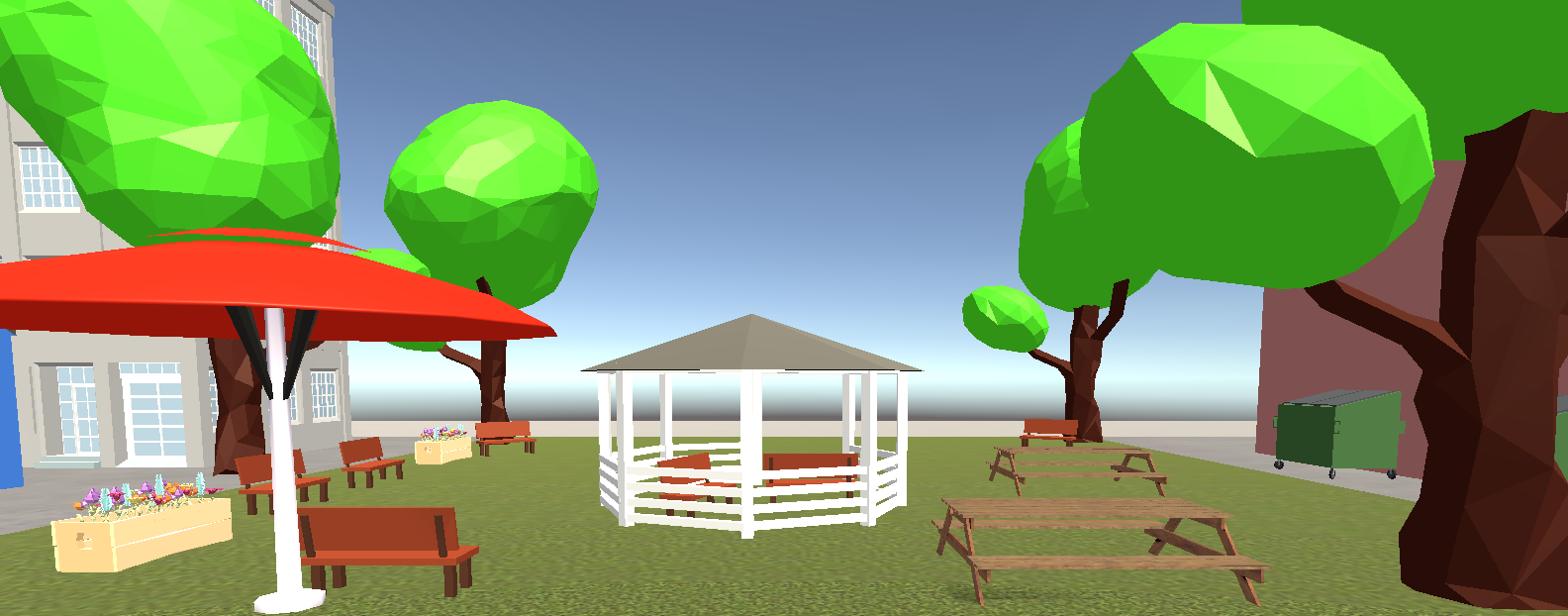}
 & 5.00          & 4.45                   & \textit{\textbf{5.18}} & 4.82          & 4.74                   & 4.67          & 4.76          & 5.13                   \\ \hline
A2 &\includegraphics[width=1.25in]{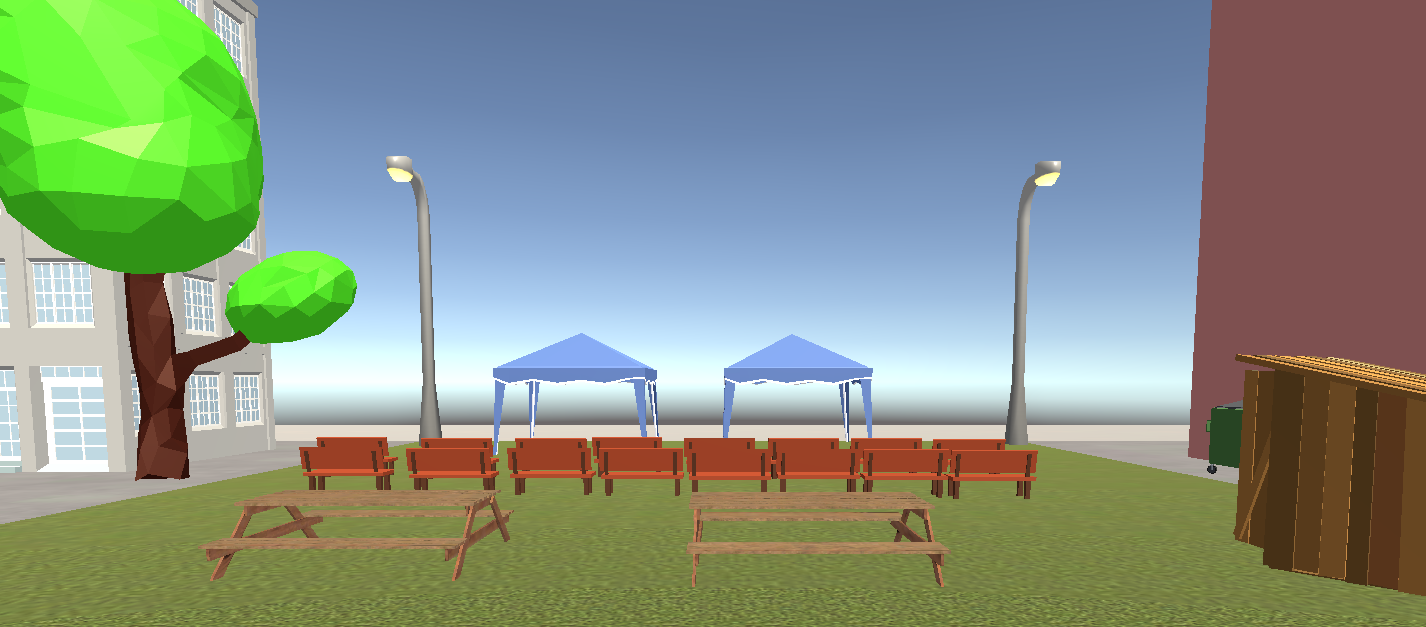} & 4.46          & 4.18                   & 5.14                   & 4.93          & 4.49                   & 4.74          & \textit{5.12} & \textbf{5.36}          \\ \hline
A3 &\includegraphics[width=1.25in]{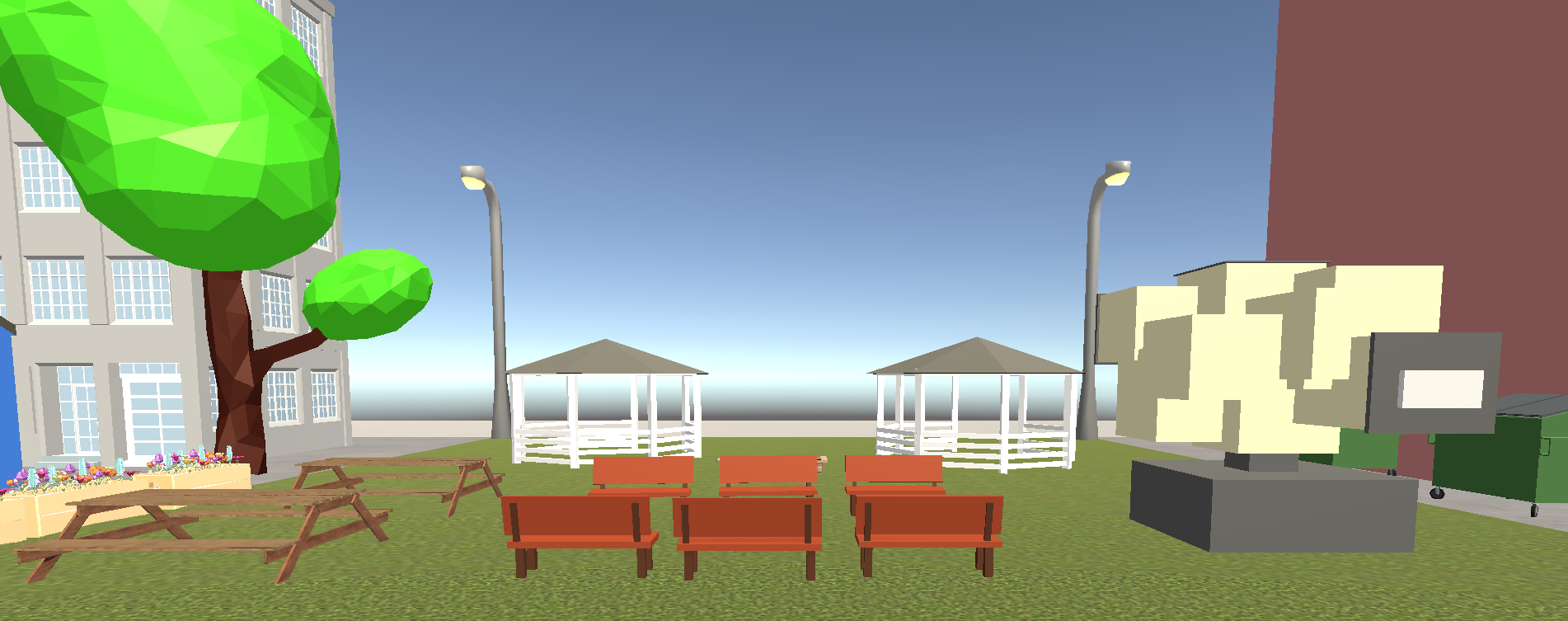} & 4.88          & 4.47                   & \textit{5.13}          & 4.79          & 4.61                   & 4.70          & 4.89          & \textit{\textbf{5.56}} \\ \hline
A4 &\includegraphics[width=1.25in]{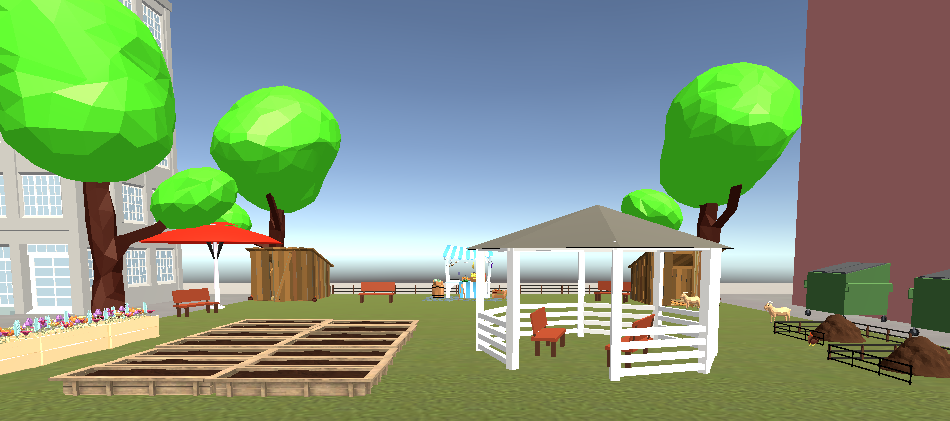} & 4.50          & 4.30                   & 4.47                   & 4.52          & \textit{\textbf{5.02}} & 4.67          & 4.42          & \textit{\textbf{5.02}} \\ \hline
B1 & \includegraphics[width=1.25in]{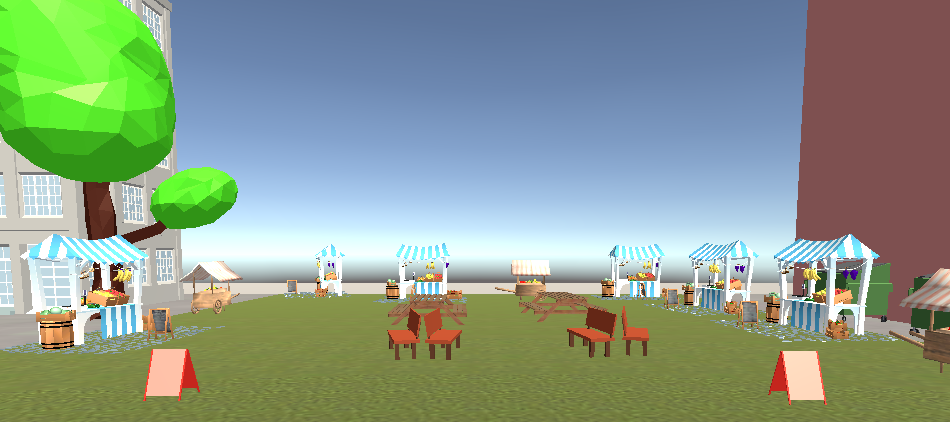} & 4.88          & 4.39                   & 4.93                   & 4.87          & 4.65                   & \textit{4.77} & 5.06          & \textit{\textbf{5.38}} \\ \hline
B2 & \includegraphics[width=1.25in]{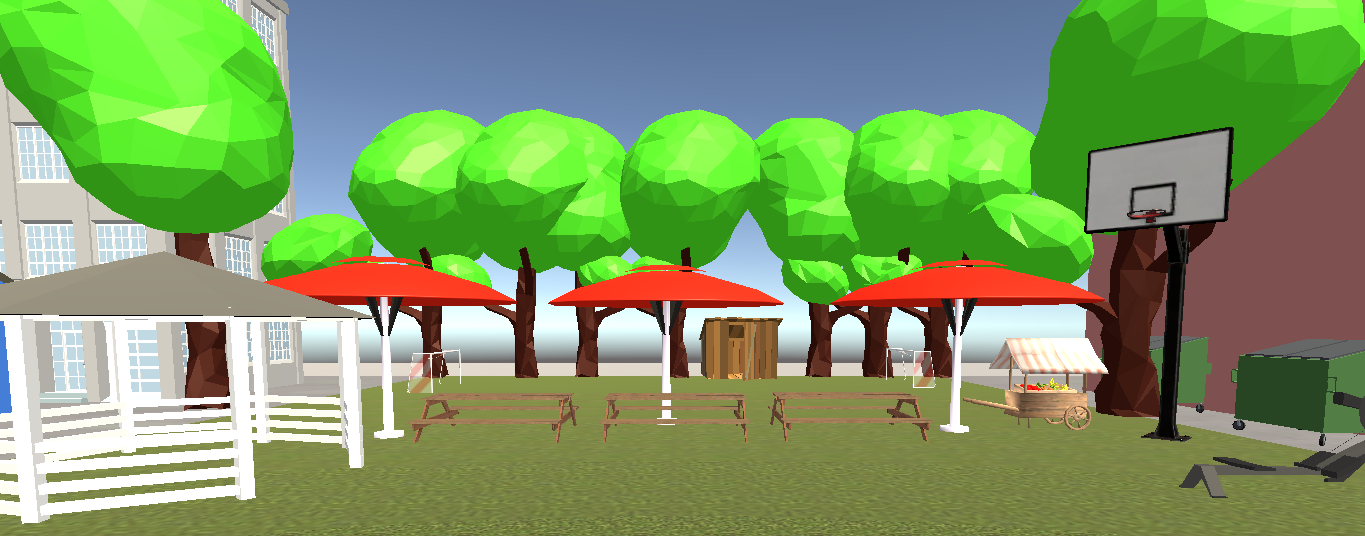} & 4.49          & 5.14                   & 4.67                   & 4.92          & 4.66                   & 5.15          & 4.92          & \textit{\textbf{5.38}} \\ \hline
B3 & \includegraphics[width=1.25in]{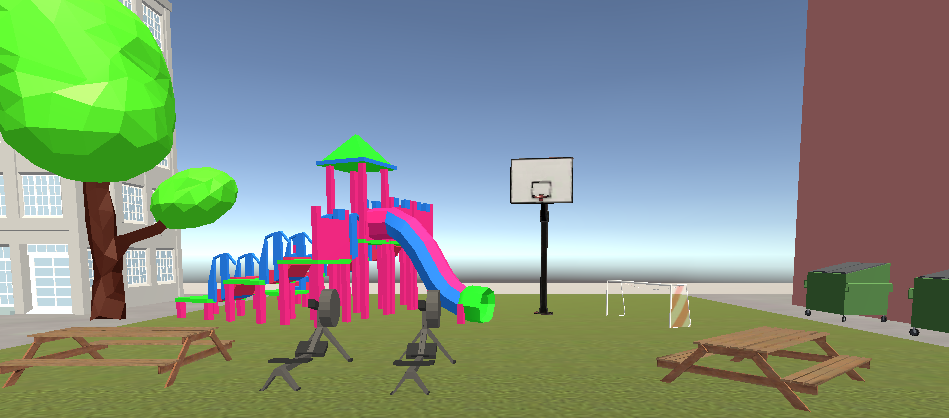} & 4.70          & \textit{\textbf{5.54}} & 5.07                   & \textit{5.02} & 4.68                   & 4.97          & 4.66          & 5.19                   \\ \hline
B4 & \includegraphics[width=1.25in]{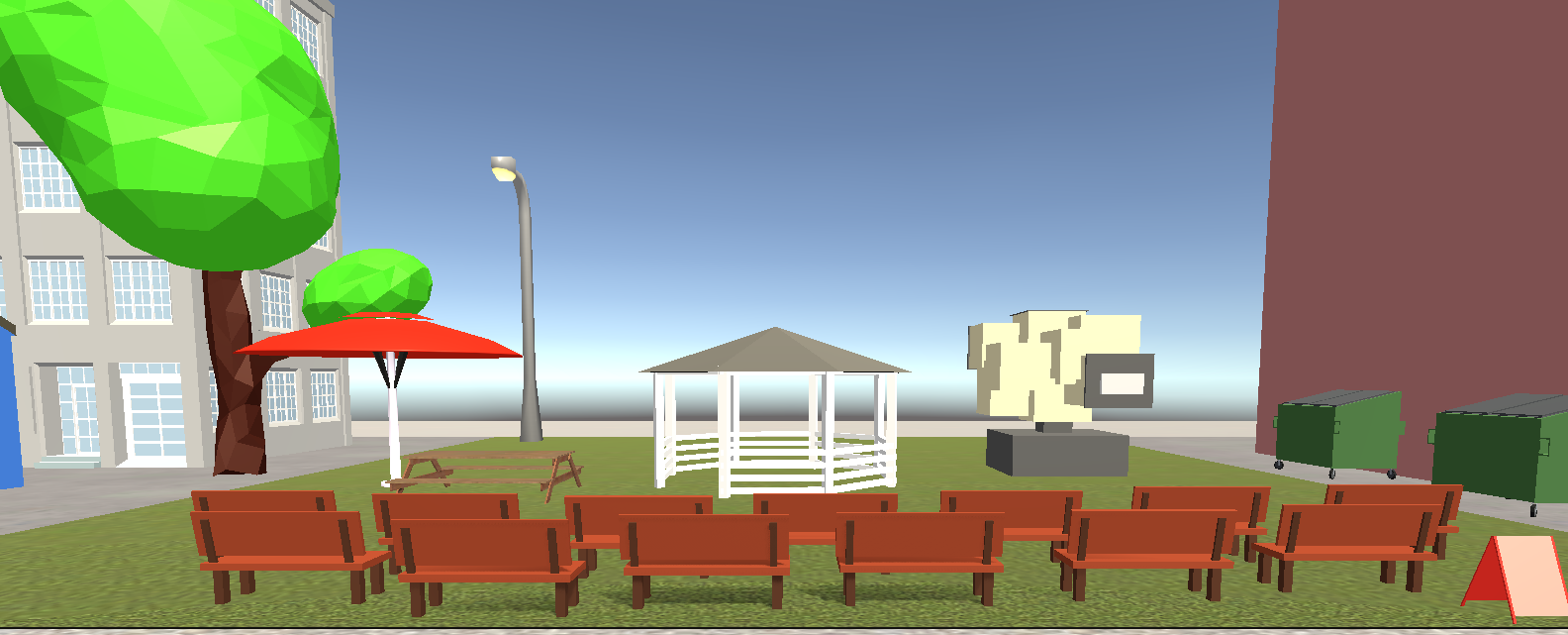} & 4.47          & 4.35                   & \textbf{5.37}          & 4.92          & 4.58                   & 4.53          & \textit{5.09} & \textit{\textbf{5.37}} \\ \hline
C1 & \includegraphics[width=1.25in]{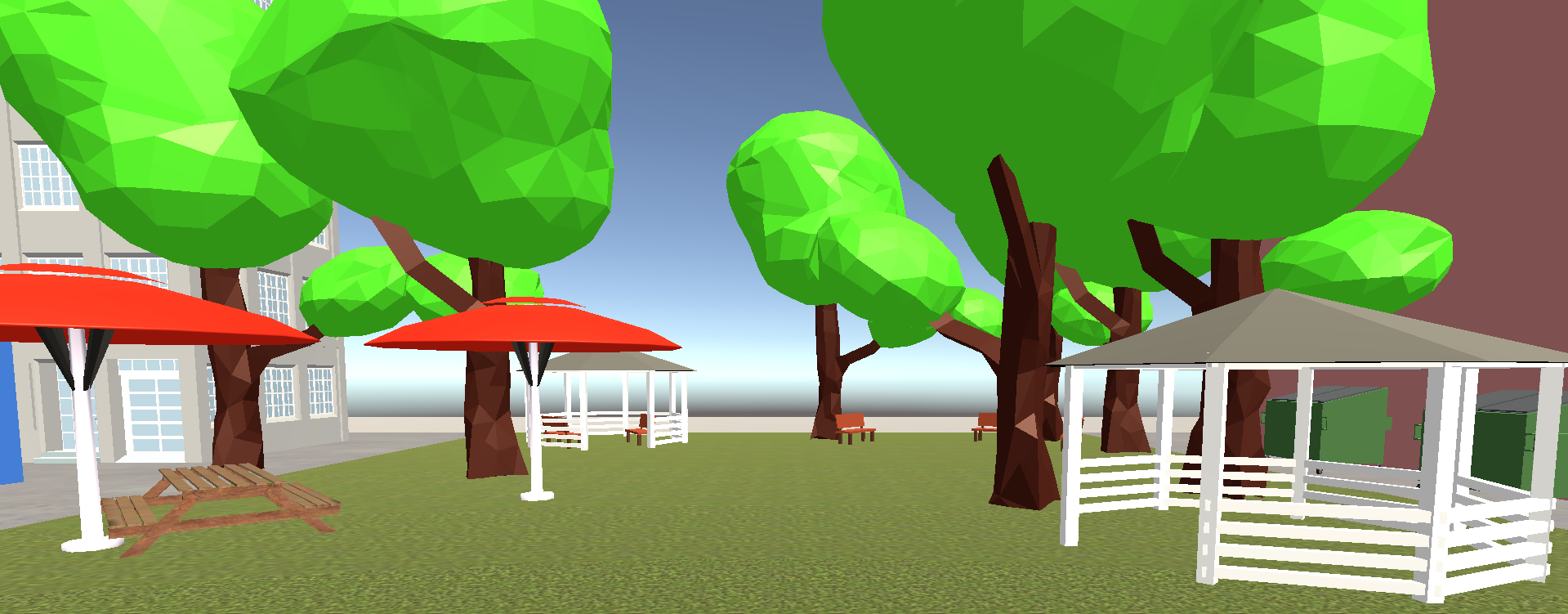} & \textit{5.15} & 4.69                   & \textit{5.37}          & 4.93          & 5.11                   & 4.77          & 4.95          & \textbf{5.47}          \\ \hline
C2 & \includegraphics[width=1.25in]{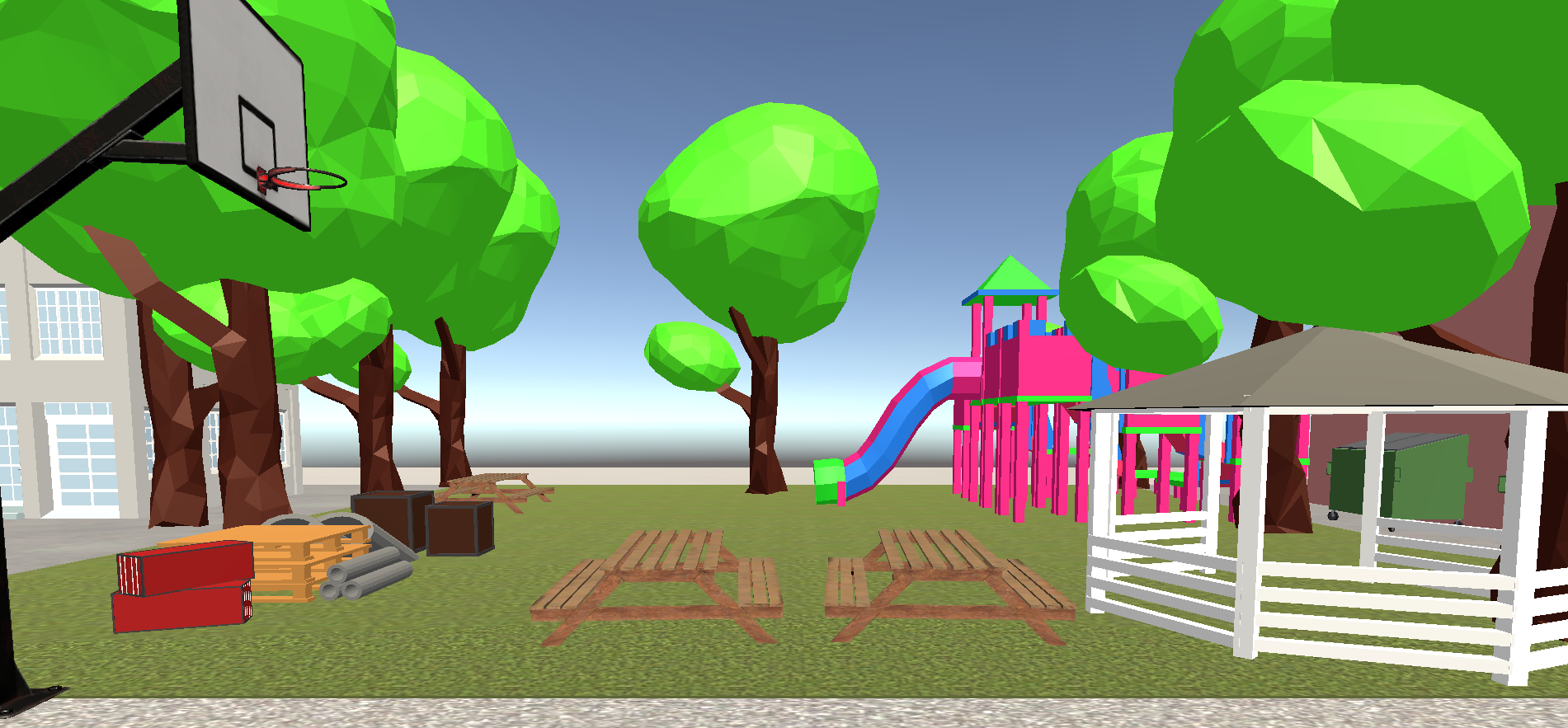} & \textit{4.70} & \textit{\textbf{5.82}} & 5.03                   & \textit{5.13} & 4.90                   & 4.90          & 4.65          & 5.41                   \\ \hline
C3 & \includegraphics[width=1.25in]{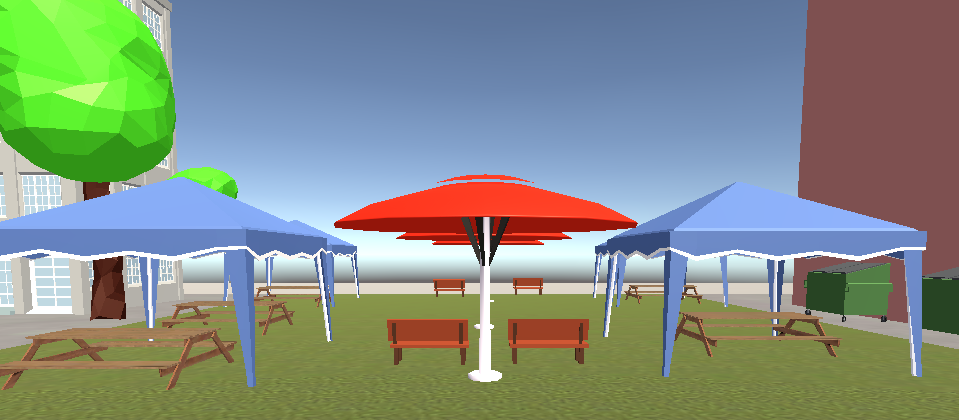} & 5.14          & 4.56                   &\textit{5.44}                   & \textit{4.92}          & 5.00                   & 4.74          & 4.92          & \textit{\textbf{5.62}}          \\ \hline
C4 & \includegraphics[width=1.25in]{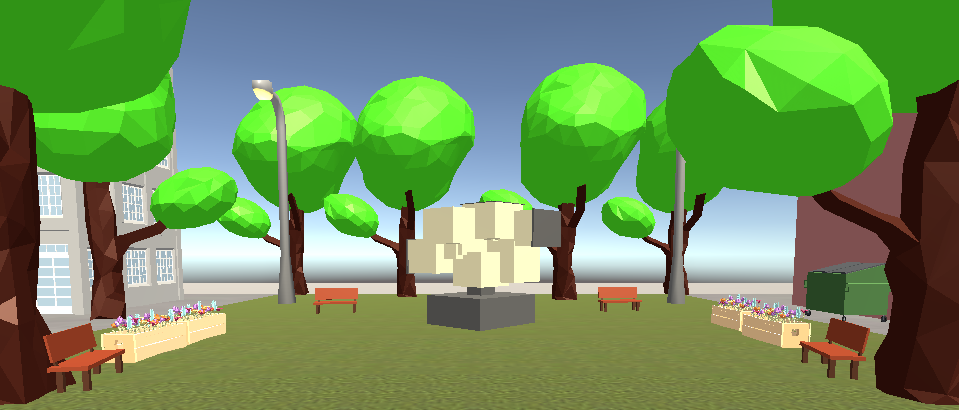} & 4.72          & 4.55                   & \textit{5.12}          & 4.77          & \textit{4.71}          & 4.94          & 4.86          & \textbf{5.26}          \\ \hline
\end{tabular}
\caption{The average rating (out of 7) for each metric for each of the 12 scenarios. \textbf{Bold} denotes the metric with the highest average.  \textit{Italics} denotes the metrics the research team chose as most representative for the scenario.}
\label{tab:averages}
\end{table*}

In nine of twelve cases, the metric with the highest average rating lines up with a metric the research team felt was most representative of the scenario. The three exceptions are A2, C1, and C4, which all had sociability as the highest-rated metric. In all three of these cases, the metrics the research team felt were most representative of the scenario were rated among the top three. Thus, we feel that users were able to communicate the essence of the scenarios through the designs they created on PatternPainter.  

It is worth noting that for nine of twelve scenarios sociability was the most highly rated metric, and no scenario had an average rating less than five on sociability. We feel that this tracks with our goals for the PatternPainter system.  While the specific use case for the space is varied across scenarios, all of them are intended as some kind of community gathering space, and sociability captures this general purpose, even if it does not capture the specific use case. 

It is also notable that across the entire table, the highest average rating is 5.82 of 7, while the lowest is 4.18. This indicates that all eight characteristics represent most of the scenarios to some degree. 

In the next section, we analyze the qualitative responses to get a better idea of which scenarios were communicated most effectively, and explore other themes that emerged in the responses.

\subsection{Qualitative Analysis}
The qualitative data was gathered by asking participants to describe each design in terms of use (whom and for what) and purpose. Due to issues with language fluency of the participants, we did not analyze all of the qualitative data.  The data was reviewed by a member of the research team, and was retained for analysis if the response sufficiently answered the questions posed and could be understood by a native English speaker with minimal effort to interpret odd or incorrect grammatical structures.  There were three common response types that were discarded: 1) single words or very short answers, such as ``park'' or ``children playing'', 2) lists of one or more of the eight metrics, and 3) hard to decipher grammar, such as, ``The place is park reception arrangement of people of this place. uses of peoples. the place is very nice.''  We retained 1-3 responses for 27 of the 28 surveys, resulting in a total of 47 responses per scenario. 

To analyze the qualitative data, we marked those responses which directly or indirectly captured elements of the scenario.  For example, a response directly capturing scenario B1 would use the words ``farmer's market'' whereas words like ``buy'' and ``sell'' would be considered capturing it indirectly. We also looked for other repeated themes, in particular where there was a different use repeatedly mentioned for a scenario.

\begin{figure}[t!]
\includegraphics[width=0.95\linewidth]{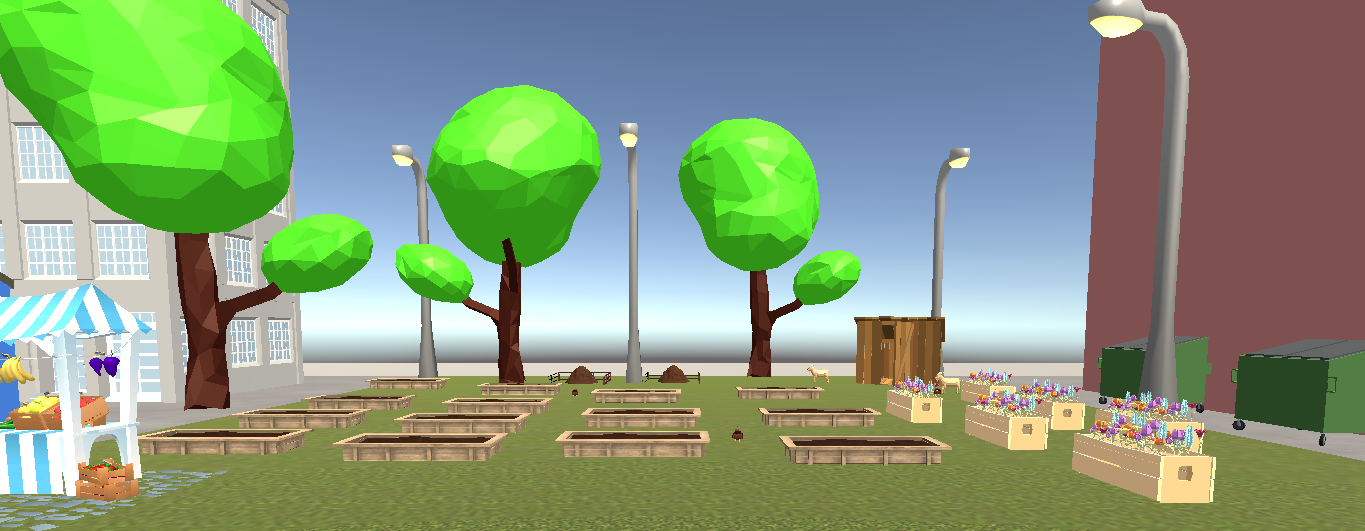} \\
\vspace{2mm}
\includegraphics[width=0.95\linewidth]{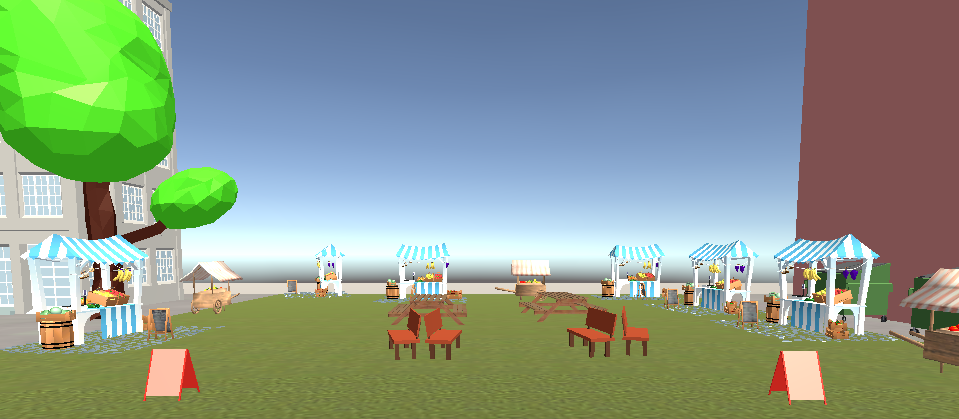}
\\ \vspace{2mm}
 \includegraphics[width=0.95\linewidth]{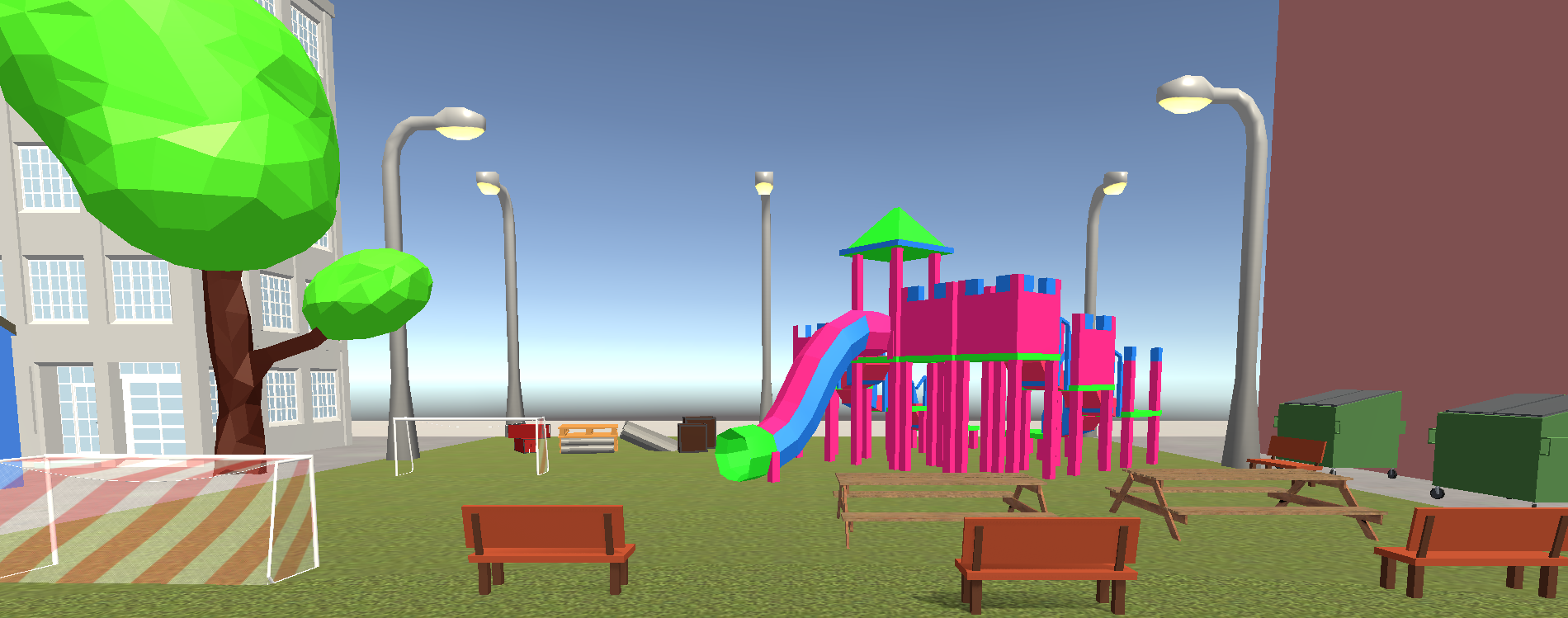}
    \caption{From top to bottom: Examples of designs for scenarios A4, B1, and C2, which successfully conveyed the purpose and use to survey respondents.}
   \label{fig:success}
\end{figure}

We found that some scenarios were communicated very successfully, a few unsuccessfully, and the rest in between.  Those that were communicated most successfully were A4, B1, and C2.  For A4, 20 of 47 responses directly mentioned the phrase ``community garden'' while another 8 used words like ``growing'' and ``planting.''  In the case of B1, 19 responses used the phrase ``farmers market'' and another 16 mentioned ``vending,'' ``selling,'' or referenced a generic community market. For scenario C2, 12 responses indicated the space was for children without mention of families or parents, and 26 directly mentioned either families or parents and children. We believe that these scenarios were most successful due to highly recognizable elements associated with each scenario---garden and flowerbeds for A2, market stalls and food carts for B1, and a playground for C1.  Figure~\ref{fig:success} shows example designs for each of these scenarios featuring these items.  

In contrast, we found that those scenarios that were the least successful were those with a very specific purpose or scenario, but without a specific set of highly recognizable and related elements. These were A1, B3, C3, and C4. While A1 was rated highly on comfort and sociability, which we believe would appeal to an elderly population, no response specifically mentioned this demographic. In contrast to a jungle gym, which is clearly intended for children, there is no analogous item that clearly signifies the elderly. The designs for B3 seemed to convey that the space was meant for children, and a few responses mentioned exercise, but the specific idea of intentionally promoting healthy habits was lost.  Scenario C3 was largely seen as a picnic or dining area due to widespread use of picnic tables in many of the designs.  However, knowing the intended purpose, it is easy to see how children might gather at these tables to study. No one captured the intended purpose of the spaces designed for C4, mainly surmising it was a space for relaxation or art exhibits, due to frequent use of benches and the statue element.  

From these less successful scenarios, it is clear that some purposes simply need context, but we do not feel that this undermines PatternPainter's usefulness.  In a real-world use case, a description of the intended purpose for the revitalized space would almost certainly be included with the designs. Due to space and attention constraints, we only presented one view of each design in the survey. Ideally users would show off a variety of angles or a allow 3D interaction with their design, and would have a written description or be there to explain the concept in person.

There is also some question as to how the elements chosen for the software affected the designs.  We consider scenario A2 as an example. While many responses captured the general intention of an entertainment space for scenario A2, we might consider whether the designs would have been more successful had we included a stage as opposed to the tents and gazebos used to create a makeshift stage area in many designs, including the one shown in Figure~\ref{fig:teaser}. We discuss this issue in more detail in the next section, where we consider areas for improvement and expansion of PatternPainter.  

\begin{figure}[]
\includegraphics[width=0.95\linewidth]{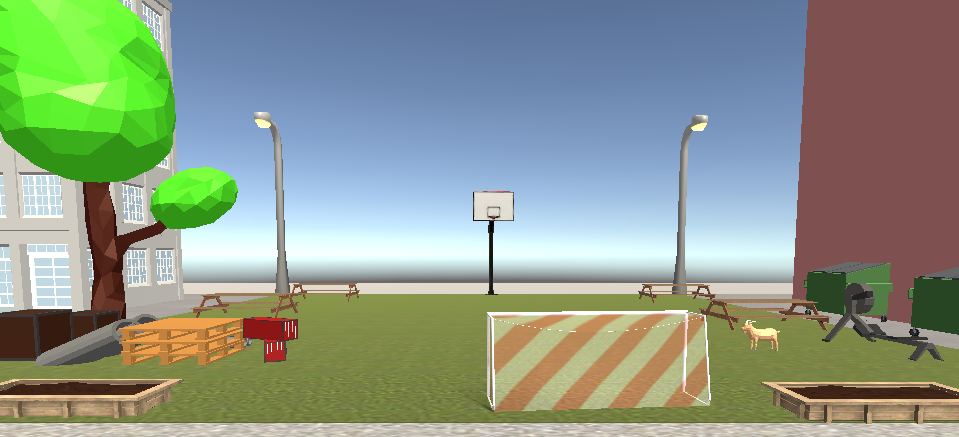} \\ \vspace{2mm}
\includegraphics[width=0.95\linewidth]{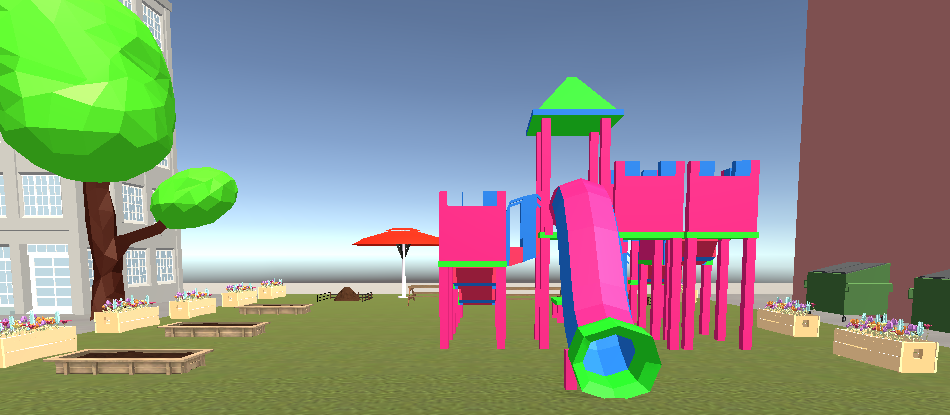}
\\ \vspace{2mm}
 \includegraphics[width=0.95\linewidth]{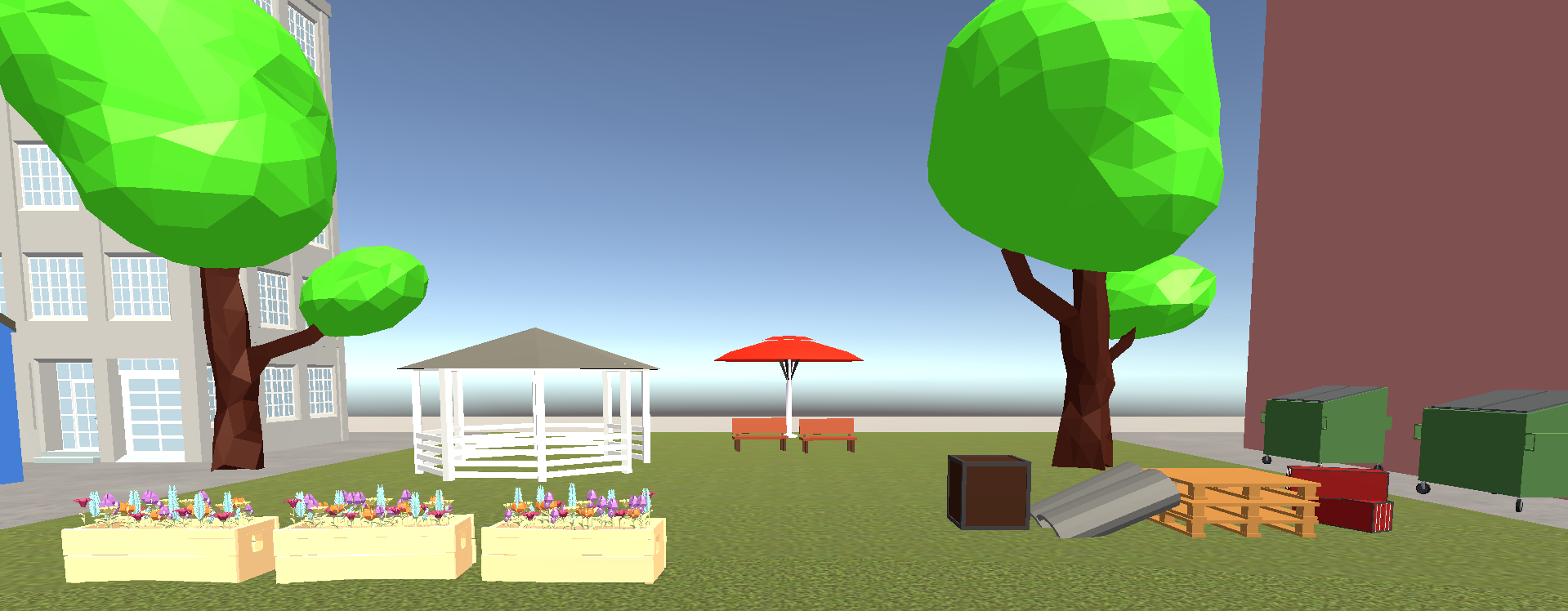} 
 \\ \vspace{2mm}
 \includegraphics[width=0.95\linewidth]{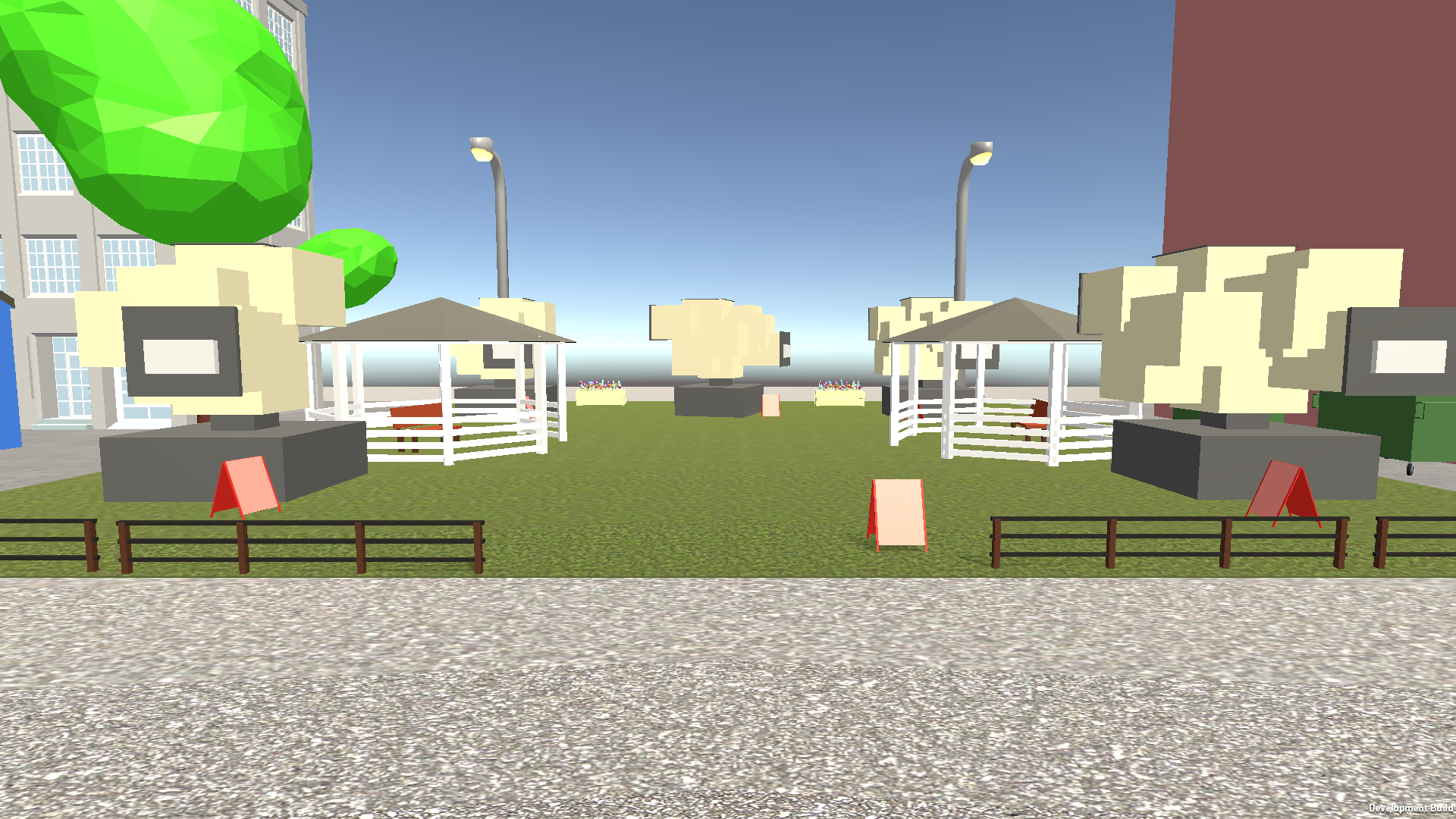}
    \caption{From top to bottom, the models which were mistaken for other things: goat (far right side) as a dog, garden beds (far left side) as sandboxes, miniature adventure park (right side) as a skate ramp, and statues (throughout) were simply unclear.}
 \label{fig:mistake}
\end{figure}

Another key theme that emerged from the qualitative analysis was that some of the elements were mistaken for other things. The goat was mistaken as a dog, the garden plots for sandboxes, and what was intended to be a miniature adventure park (see pattern \#73 \cite{alexander1977pattern}) was mistaken for a skatepark by five respondents, and a dangerous one at that, as one respondent noted, ``Those are probably dangerous though as they seem unfixed.'' Several respondents were simply unclear about the statue element referring to them as, ``the blocky things'' and ``THOSE MINECRAFT SHEEP STATUE THINGS.'' Figure~\ref{fig:mistake} shows these four items in the context where they were mistaken for these other things.

In these instances, the unclear 3D models may have failed to communicate the correct context for the scenario. For instance, the second image in Figure~\ref{fig:mistake} is meant to be a community garden (scenario A4), but the combination of a playground with the garden beds caused them to look like sandboxes, erasing the context of a garden.  However, as mentioned previously, in real-world use cases context would be provided with designs to help mitigate such issues. Furthermore, having some models that are flexible in their use is not inherently bad, as it broadens the scope of objects available to designers.   

Based on these experiments, we feel that PatternPainter was generally successful in helping ordinary people create and communicate designs for re-purposing an urban lot. However, there are certainly areas for future work and improvement, which we discuss in the next section.  

\section{Discussion}
In this section, we first discuss several areas for improvement an future work based on our experimental results and some feedback given to us by Chris Tallman, an expert designer with extensive experience in participatory design for urban planning. We then summarize the lessons we learned building and testing the PatternPainter system into three general design goals for technological aids that allow ordinary citizens to design their own tactical, urban revitalization projects.  

\subsection{Future Work}
As mentioned in the previous section, one of the major questions with a system like PatternPainter is what elements to include. We attempted to provide a sufficient array of elements to fit each of the pre-defined scenarios, but in the future users may want to use PatternPainter to brainstorm without a clear use case in mind. While we used Alexander's patterns as inspiration for the scenarios and elements, as Chris Tallman noted, ``I was surprised at both how closely Alexander and company identified the armature of whole landscape patterns but more so by how many are missing.'' He then asked, ``What order of complexity is there to having a tool where the user is walked through defining their own patterns?''~\cite{Tallman}.

We feel that going beyond Alexander's language to capture more local knowledge as well as to solve problems that have cropped up in the almost 40 years since the book's 1977 publication is an important extension of the work. For instance, the disruption of public education due to the 2020 COVID-19 crisis has shown widespread inequalities in access to broadband Internet, with many students unable to access online learning tools. This might lead to a new pattern: ``Public Internet Access'' that calls for public WiFi hotspots covering a city or region, and spaces to gather to safely use this infrastructure, so that all students can connect to online learning opportunities. We can only begin to imagine what myriad other patterns communities might define based on their unique circumstances and cultures. 

 However, this brings up the related question of how to scale and support such a system.  Our first step is to open source the system, which we intend to do with PatternPainter.  This does not solve all the problems associated with scaling and maintaining this kind of system, but it is an enabler of further refinement and also helps the system to stay free.

Another suggestion of Tallman's was the inclusion of action items.  He suggests thinking about the question, ``What actions can you take today?''  He proposes comparing the design with a database of tactical actions, and then listing suggestions that can be taken quickly and easily by community members.  
We think this idea is deep and empowering, as it is a first step toward activating community members to take on the next two phases of the design thinking process---prototype and test. This is the process by which crosswalks get painted, community gardens get planted, and neighbors become friends. 

The idea of incorporating action items also alludes to the issue of creating sustained engagement in the projects designed by PatternPainter.  As Tallman notes, ``There are a vast number of popup community gardens laying fallow.'' Sustaining community engagement in local projects is an issue that has previously been studied in the context of HCI \cite{slingerland2019join}, and a problem we are also interested in addressing in future work.  However, addressing it goes beyond the scope of this particular paper.

\subsection{Implications for Design}
Based on our experience designing and evaluating PatternPainter, as well as our discussions with Chris Tallman, we came up with three general implications for design for community-led design systems for urban revitalization, which we frame as goals. The first goal is: \\

\noindent \emph{\textbf{DG1--Expertise.} Fill the gap of design expertise for ordinary community members tackling urban revitalization projects when professional design services are unavailable or impractical.}\\

We feel this kind of system should be based on expert design knowledge in some form, rendering it more than just a hodge-podge collection of elements to be strewn about a space.  We used Alexander's pattern language as the basis for our scenarios and elements, however there are many other expert works that could be substituted or included. Another area we are interested in exploring in the future is the inclusion of artificial intelligence methods to create co-creative systems that guide non-expert users in real-time based on expert design principles and knowledge. 

Once a method of filling the design expertise gap has been identified, the question becomes how to disseminate the designs. During a traditional participatory workshop, designers might lead community members to produce abstract representations such as the one seen in Figure~\ref{fig:HooverCircle}. However, while it might be possible for participants to understand plans like this one, for other community members it will likely be much more difficult to visualize the redesigned space. Our goal is to make it easy for all community members to imagine proposed changes to a space, leading to our second design goal:\\

\noindent \emph{\textbf{DG2--Visualization.} Enable ordinary community members to output expert designs in an intuitive and easy to view format.} \\

For the purpose of PatternPainter, we used 3D visualization, but know there may be other suitable methods, including augmented reality, which we hope to explore in the future. 

Finally, not only is our goal to help community members imagine proposed upgrades with intuitive visualization, but also to help them think about pushing the boundaries of what is possible.  For example, when designing a park, it might be easy to imagine that trees are good for their provision of shade, air filtration, and natural habitat.  However, it may be the case, for example, that \textit{fruit} trees are better in specific settings; not only do they provide the benefits of trees in general, but they also serve as a local food source.  In this spirit, our third design goal is:\\

\noindent \emph{\textbf{DG3--Imagination.} Help ordinary community members stretch their imaginations to consider new and nontraditional uses for urban space.} \\

We took a small step toward this in PatternPainter by trying to include some elements that are not commonly considered part of the urban landscape, such as goats, chickens, and compost piles, but we still have a great deal of work to do to toward achieving this goal. 

\section{Conclusion}
Leaving city planning to governments (particularly in the US context) has yielded only crumbling infrastructure (in 2017 the American Society of Civil Engineers gave the US a D+ for infrastructure~\cite{infr_report}), slow and unreliable public transit~\cite{transit_17,transit_18,graehler2019understanding}, and a dearth of green space, particularly in areas of lower socioeconomic status~\cite{heynen2006political,wolch2014urban}.  We believe it is time to put city planning and urban repair back into the hands of the people of each neighborhood. The blue-collar bus-rider should dictate the bus schedule, not the transit director who drives his SUV to work; the mother and child navigating broken swings and unshaded park benches should design the parks, not consultants flown in from out of state; and the urban gardener with no yard should be free to plant community food forests rather than leaving blighted lots behind the fences of a city's public works department. PatternPainter is a first step toward helping citizens take back the power for planning and repairing their communities. Based on the guiding principles of tactical urbanism \cite{lydon:2015:tactical} and \textit{A Pattern Language} \cite{alexander1977pattern}, and based on our experiments has shown great promise in helping ordinary people create and communicate deigns for urban revitalization projects. 

Our expert correspondent, Chris Tallman, responded positively to the PatternPainter prototype, and suggested a few features to further improve the tool. Based on these suggestions and our experimental evaluation, our aim for the near future of PatternPainter is to modularize the system to enable the open-source community to contribute modules for additional patterns, to integrate GIS to allow for location-specific plans, and to allow for other types of urban repair. We are also looking to design tools to assist in other phases of the design process.

\bibliographystyle{ACM-Reference-Format}
\bibliography{references}


\end{document}